\documentclass[12pt]{article}
\usepackage{graphicx}
\usepackage{epsf}

\def\dst{\displaystyle}
\def\f{\frac}
\def\b{\beta}
\def\x{\xi}
\def\vp{\varphi}

\def\m{\mu}
\def\s{\sigma}
\def\t{\theta}
\def\p{\partial}
\def\ep{\epsilon}
\def\e{\varepsilon}

\def\eps{\varepsilon}
\def\be{\begin{equation}}
\def\ee{\end{equation}}
\def\bea{\begin{eqnarray}}
\def\eea{\end{eqnarray}}
\def\a{\alpha}
\def\ba{\begin{array}}
\def\ea{\end{array}}

\def\l{\left}
\def\r{\right}
\def\Det{\mathrm{Det}}
\def\Tr{\mathrm{Tr}}
\def\Li{\mathrm{Li}}

\def\gsim{\raisebox{-0.6ex}{$\stackrel{\textstyle >}{\sim}$}}

\begin{document}
\begin{titlepage}
\begin{center}
{\Large \bf William I. Fine Theoretical Physics Institute \\
University of Minnesota \\}
\end{center}
%\vspace{0.2in}
\begin{flushright}
FTPI-MINN-09/15 \\
UMN-TH-2743/09 \\
April 2009 \\
\end{flushright}
%\vspace{0.3in}
\begin{center}
{\Large \bf  Spontaneous and Induced Decay of Metastable Strings and Domain Walls.
\\}
\vspace{0.2in}
{\bf A. Monin \\}
School of Physics and Astronomy, University of Minnesota, \\ Minneapolis, MN
55455, USA, \\
and \\
{\bf M.B. Voloshin  \\ }
William I. Fine Theoretical Physics Institute, University of
Minnesota,\\ Minneapolis, MN 55455, USA \\
and \\
Institute of Theoretical and Experimental Physics, Moscow, 117218, Russia
%\\%[0.2in]
\end{center}

\begin{abstract}
We consider decay of metastable topological configurations such as strings and domain walls. The transition from a state with higher energy density to a state with lower one proceeds through quantum tunneling or through thermally catalyzed quantum tunneling (at sufficiently small temperatures). The transition rate is calculated at zero temperature including the preexponential factor and also at a finite low temperature.  The thermal catalysis factor is closely related to the probability (effective length) of destruction of the string (the domain wall) in collisions of the Goldstone bosons, corresponding to transverse waves on the string (wall). We derive a general formula which allows to find the probability (effective length) of a string (wall) breakup by a collision of arbitrary number of the bosons. We find that the destruction of a string only takes place in collisions of even number of the bosons, while the destruction of the wall can occur in a collision of any number of particles. We explicitly calculate the energy dependence of such processes in two-particle collisions for arbitrary relation between the energy and the largest infrared scale (the size of a critical gap).
\end{abstract}

\end{titlepage}
\tableofcontents

%\newpage

\section{Introduction}
Topological configurations of fields with the geometry of a string or a domain wall arise 
in various models either as solutions of classical field equations or as nonperturbative effective configurations of essentially quantum fields.   String-like configurations are present in polymers, superconductors, in theoretical models of non-Abelian dynamics, as well as in models of QCD confinement, and as topological defects in models with spontaneous breaking of gauge or global symmetries. Metastable domain wall solutions arise in models with spontaneously broken approximate symmetry. The existence of such solutions was shown from different points of view, for example, in Refs.~\cite{shifman97,witten98a,witten98b}.

\begin{figure}[ht]
  \begin{center}
    \leavevmode
    \epsfxsize=5cm
    \epsfbox{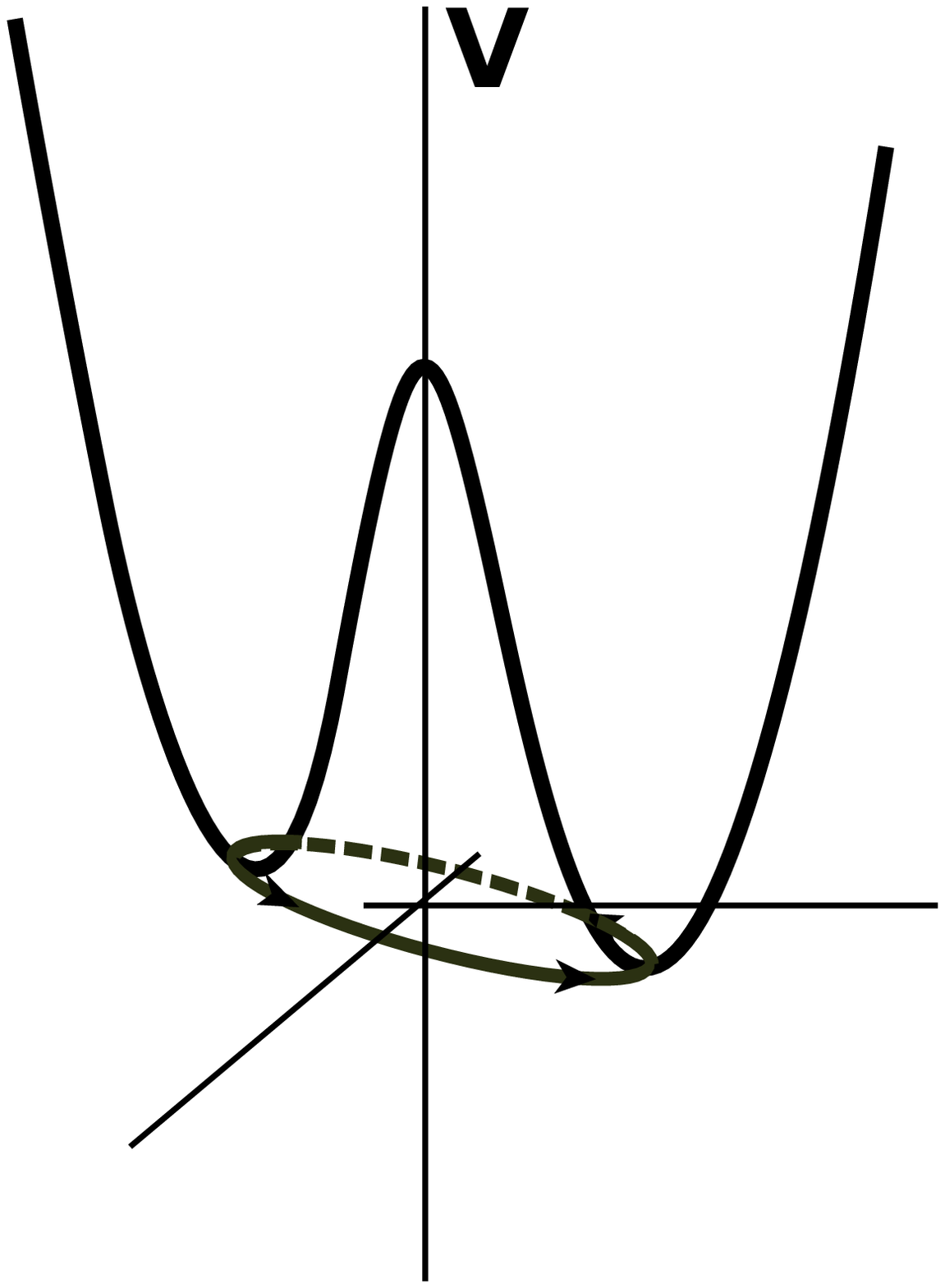}
    \caption{Potential}
  \end{center}
\label{potential}
\end{figure}

These configurations are classically stable, since there is no classical trajectory which takes from one topological solution into another with different topological number. But it may be the case that a classically forbidden trajectory for such a deformation exists so that the considered configurations are unstable quasiclassically and decay due to e.g. quantum tunneling effects. As an illustration of such decay process one can consider the example of a metastable domain wall investigated in Refs.~\cite{fzh,mv08w} which arises in a model with a scalar field potential shown in Fig.\ref{potential}. The domain wall solution corresponds to the interpolation between the same vacuum state at two spatial infinities, e.g. at $z=-\infty$ and $z=+\infty$ with the field winding around the `peg' in the potential.
Such configuration is classically stable in the sense that the solution with
certain winding number can not evolve classically to a
solution with different topological number. However, such a transition can
proceed either due to a thermal fluctuation, where the path, corresponding to the
solution, is lifted over the barrier or due to quantum tunneling (the motion
in a classically forbidden region).

The topological configurations can be metastable with respect to either a complete breaking, or a transition to an object of lower tension emerging instead of the initial one. The former situation is relevant e.g.
for a break up of a QCD string with formation of a quark - antiquark pair, or a
formation of a monopole - antimonopole pair\cite{Vilenkin}, and also in a whole
class of theories with spontaneous symmetry breaking\cite{Preskill}. The latter
situation involving a phase transition between states of a string with different
tension is found e.g. in Abelian Higgs models embedded in non-Abelian
theories \cite{Shifman,mv08s}.

A domain wall decay is analogous to the well-known spontaneous false vacuum decay process \cite{vko,Coleman,mv2004} in $2+1$ dimensions. Indeed, if a hole of area $ \mathcal {A} _ c $ is created in a wall with tension $\ep$, the gain in the energy is $\ep \, \mathcal {A} _ c $. The barrier that inhibits the process is created by the energy $\s \mathcal{P}$, with $\mathcal{P}$ being the perimeter of the hole and $\s$ being a tension associated
with the interface. Thus, the `area' energy gain exceeds the barrier energy only
starting from a critical size of the hole created, i.e. starting with a round
hole of radius $R _ w = 2 \s/\ep$. Once a critical bubble of the lower phase has nucleated due to
tunneling, it expands, converting the domain wall. Therefore the probability of
the transition is given by the rate of nucleation of the critical holes in the
domain wall.

The same argument also applies to the string decay which is similar to a metastable vacuum decay process in $1+1$ dimensions. The critical size of the gap in the initial string is then $ \ell _ c = 2 R _ s = 2 \, \m / \e$. Moreover due to the geometry of the problem such decay also bears a great similarity to the well known Schwinger process of  production of pair of charged particles by an external electromagnetic field \cite{Schwinger}.

Despite the similarity with the false vacuum decay the breaking of the strings and walls involves an essential difference associated with the transverse waves propagating on the latter objects, corresponding to massless bosons, which in fact are the Goldstone bosons of the spontaneously broken translational invariance. The effect of the soft modes of the field of these bosons enters the probability of breaking the topological defects at the preexponential level~\cite{mv08w,mv08s}.

The process of a metastable string transition from a state with tension $\e_1$ to a state with tension $\e_2$ was considered in \cite{Vilenkin, Preskill, Shifman}. Using the analogy with the false vacuum decay the decay rate was found with exponential accuracy in terms of the rate of nucleation of critical gaps per unit length of the string, $\gamma_s = {d\Gamma / d \ell}$\,\footnote{It should be mentioned that in $d=2$ there is no transverse motion for the string, and the preexponential factor can be also copied from the known result of spontaneous metastable vacuum decay 
%\be
$$ C_{d=2}= {\eps_1-\eps_2 \over 2 \pi}~.$$
%\label{c2}
%\ee
},
\be
\gamma_s =C \, \exp \left ( - \, {\pi \, \mu^2 \over
\eps_1-\eps_2} \right )~,
\label{g0}
\ee
where $\m$ is the mass associated with the interface between the two phases of the string.
The decay rate of an axion domain wall was considered at the level of the semiclassical exponent in Ref.~\cite{fzh} by adapting
the expression for metastable vacuum decay in $2+1$ dimensions.

The preexponential factors for the spontaneous decay of string and walls were calculated in \cite{mv08w,mv08s}. For a string the rate of the transition between the states with tension $\e_1$ and $\e_2$ was found to be
\be
\gamma_s = {\eps_1-\eps_2 \over 2 \pi} \, \left [ F \left ( {\eps_2
\over \eps_1} \right ) \right ]^{d-2} \, \exp \left ( -
\, {\pi \, \mu_R^2 \over \eps_1-\eps_2} \right )~,
\label{gf}
\ee
with the dimensionless factor $F$ given by
\be
F \left ( {\eps_2 \over \eps_1} \right ) = \sqrt{\eps_1+\eps_2 \over 2 \,
\eps_1} \,\, {\rm \Gamma} \! \left
({\eps_1+\eps_2 \over  \eps_1 - \eps_2} +1 \right ) \, \left ( {\eps_1-\eps_2
\over  \eps_1 + \eps_2} \right )^{\eps_1+\eps_2 \over  \eps_1 - \eps_2} \, \exp
\left ( {\eps_1+\eps_2 \over  \eps_1 - \eps_2} \right )  \left ( 2 \pi \,
{\eps_1+\eps_2 \over  \eps_1 - \eps_2} \right )^{-1/2}~,
\label{ff}
\ee
while the decay rate of a domain wall is formulated in terms of the rate $\gamma_w$ of nucleation of critical holes per unit area, $\gamma_w = {d\Gamma / d A}$,
\be
\gamma_w = \f {\tilde{\mathcal {C}}} {\ep ^ {7/3}} \, \exp \l( - {16
\, \pi \, \s _ R ^ 3 \over 3 \, \ep^2} \r) ~,
\label{rate}
%\label{prevac}
\ee
where $ \m _ R $ and $ \s _ R $ are renormalized mass and tension parameters associated with the interface of the strings and domain walls respectively, and $\tilde{\mathcal {C}}$ is a constant that does not depend on $\ep$ but does depend on other dimensional parameters in the underlying field theory.

The thermal effects in the decay of metastable strings~\cite{mv2} further expose the difference from the false vacuum decay and the Schwinger process. Namely, as long as the temperature $T$ is lower
than the inverse of the critical length, $T < 1/\ell_c$ the thermal effects in
the latter processes are exponentially suppressed as $\exp(-m/T)$ with $m$ being
the lowest scale for particle masses in the theory, and these effects are
very  small due to the strongly suppressed presence of massive particles in
the thermal equilibrium\cite{Garriga94}. In the case of a string, however, the
transverse waves on the string are massless so that their excitation has no
suppression by the mass at arbitrarily low temperature. The thermal excitations
of these waves create fluctuations in the distribution of the energy in the string
which catalyze the nucleation of the critical gap. Clearly, at $T \ll 1/\ell_c$
the typical wavelength of the thermal waves $ \sim 1/T $ is large in
comparison with the critical length $\ell_c$, and the thermal effect in the rate
is quite small, although not exponentially small. The leading low temperature
correction in the nucleation rate is given by the thermal catalysis factor\cite{mv2}
\be
\mathcal { K } _ s =  1 + (d-2) \, {\pi^8 \over 450}\, \left ( {\e_1-\e_2 \over 3 \e_1-\e_2} \right
)^2 \, \l(\f{\ell_c T}{2}\r)^8 + O\left[(\ell_c T)^{12} \right ] ~,
\label{kf}
\label{klow}
\ee
while as $T$ approaches $1/\ell_c$, the catalysis factor develops a
singularity at $\ell_c T =1$. At still higher temperatures the considered string transition behaves, in a sense, similarly to the false vacuum
decay\cite{Garriga94}, namely the regime of the transition changes to a
different tunneling trajectory, so that the temperature dependence appears in
the semiclassical exponential factor in Eq.(\ref{gf}), rather than in the
preexponential term.

The effect of the thermally excited waves in the decay rate can in fact be considered~\cite{mv09s} as an additional contribution to the probability of the decay due to collisions of the Goldstone bosons that are present in the thermal bath. As it turns out, it is possible to identify in the thermal catalysis factor the contribution of individual such collisions and thus calculate the probability of the destruction of metastable string by colliding particles. In particular, the first temperature dependent term in the expansion (\ref{klow}) is entirely due to the process of the critical gap creation in a collision of two
particles in the limit, where their center of mass energy $E=\sqrt{s}$ is much
smaller than $1/\ell_c$. In order to separate the terms in the decay probability originating from collisions of different number of the bosons, the standard thermal approach to calculating the decay rate at finite temperature~\cite{mv2} is modified~\cite{mv09s} by formally introducing a negative chemical potential for the
bosons. This allows to find the dependence of
the probability of string destruction in an $n$-boson collision for an arbitrary relation
between $\ell_c$ and the energy of the bosons. One of the results of such a consideration is that the string destruction takes place only in collisions of even number of bosons and is absent at odd $n$.  The explicit expression for the (dimensionless) probability $W_2$ of the
critical gap creation in two-boson collisions at arbitrary values of the
parameter $E \, \ell_c$  has especially simple
form for the decay of the string into `nothing' i.e. at $\e_2=0$:
\be
W_2=2\, \pi^2 \, \ell_c^2 \, \gamma_s \, \left [ I_3 \left ( {E \ell_c \over 2}
\right ) \right ]^2~,
\label{w20}
\ee
with $I_3(x)$ being the standard notation for the modified Bessel function of the third order.
This expression is applicable as long as the energy $E$ is small in comparison with the string energy scale: $E \ll \sqrt{\ep}$, which condition still allows for the parameter $E \, \ell_c$ to be large. At $E \, \ell_c \gg 1$ the energy dependent factor in Eq.(\ref{w20}) has the exponential behavior $\exp (E \ell_c )$ which matches the semiclassical approximation for the energy-dependent factor\cite{vs}, corresponding to tunneling at the energy $E$ rather than at zero energy. In this sense the discussed energy dependence of the collision-induced probability of the decay describes the onset of the semiclassical behavior.

One can also notice that the thermal factor $\mathcal {K} _s$,  is singular~\cite{mv2} at $\ell_c T = 1$. However
the two-boson production rate does not exhibit any singularity at any value of
the parameter $E \ell_c$, and a similar smooth energy behavior is also true for individual 
$n$-boson processes. Thus the `explosion' of the
thermal rate at $\ell_c T = 1$ is a result of infinite number of processes
becoming important at this point, rather than due to a finite set of processes
with a limited range of $n$ developing large probability at the energy per
particle of the order of $1/\ell_c$.

The purpose of this paper is to generalize to the case of a metastable domain wall the approach used for analyzing the thermal and collision induced effects in the decay of metastable strings. We find the general expression for the thermal catalysis factor $\mathcal {K} _w$. In particular, the leading term in this factor at low temperature (in $d$ dimensions) is given by
\be
\mathcal {K} _w = 1 + 12 \,(d-3) \, \zeta ^ 2 (5) \, (R _ w \, T) ^ {10} + \dots.
\label{kfwl}
\ee
where $R_w$ is the radius of the critical hole in the metastable wall, and $\zeta(x)$ is the standard notation for the Riemann $\zeta$ function, $\zeta(5) \approx 1.037$. We also aim at providing here a systematic and self-contained presentation of the calculations leading to our final results. To this end we recapitulate in detail in Section 2 the technically simpler calculations for the processes with metastable strings, and then in Section 3 expand the method to the case of a metastable domain wall. Such `semi-review' format of the paper also allows us to illustrate the similarity and the difference between the string and the domain wall transitions. Although the calculations in both cases are very much similar, the results differ in some essential details. In particular the destruction of a metastable wall is induced by a collision of any number of Goldstone bosons, whereas a string can be destroyed by a collision of only even number of the bosons. Another difference relates to the thermal catalysis of the decay: for a string the low temperature expansion for the enhancement of the quantum tunneling is applicable everywhere where the series is convergent, i.e. up to the temperature $T = 1/\ell_c$, while for a domain wall the thermal fluctuations start dominating\cite{Garriga94} at a lower temperature ($T = 3/(4 \ell_w)$ with $\ell_w$ being the diameter of the critical hole in the wall), at which the thermal effects in the quantum tunneling probability are still very small numerically.

%\newpage
\section{String-like configuration \label{sect1}} 
\subsection{Spontaneous decay of a metastable string}
The tunneling trajectory can be described in the Euclidean space by a
configuration called the `bounce'\cite{Coleman}, which is a solution to
(Euclidean) classical equations of motion.
The general expression for the effective Euclidean space action for the string
with the two considered phases can be written in the familiar Nambu-Goto form:
\be
S=\mu \, P + \eps_1 \, A_1 + \eps_2 \, A_2~,
\label{a0}
\ee
where $A_1$ and $A_2$ are the areas of the world sheet for the two phases, and
$P$ (the perimeter) is the length of the world line for the interface between
them.

\begin{figure}[ht]
  \begin{center}
    \leavevmode
    \epsfxsize=12cm
    \epsfbox{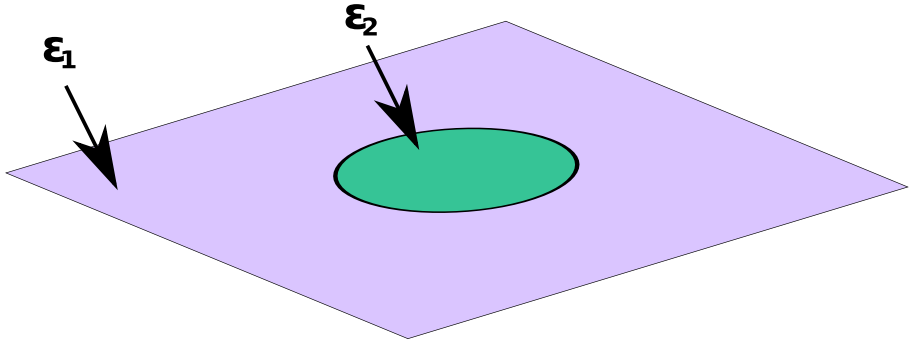}
    \caption{The bounce configuration, describing the semiclassical tunneling
trajectory.}
  \end{center}
  \label{fig2:bounce}
\end{figure}

The action (\ref{a0})  is an effective low-energy expression in the sense that
it only describes the `stringy' variables and is applicable as long as the
effects of thickness of the string and of its internal structure can be
neglected.  Denoting $M_0$ the mass scale at which such approach becomes invalid
(e.g. the thickness of the string $r_0 \sim 1/M_0$), one can write the condition
for the applicability of the effective action (\ref{a0}) in terms of the length
scale, $\ell \gg 1/M_0$, and the momentum scale $k \ll M_0$.
Assuming that the initial very long string with tension $\eps_1$ is located
along the $x$ axis, one can readily find that the action (\ref{a0}) has a
nontrivial stationary configuration, the bounce, namely, that of a disk in the
(t,x) plane occupied by the phase $2$, as shown in Figure 2, with the radius
\be
R _ s = {\mu \over  \eps_1-\eps_2}~,
\label{rc}
\ee
which is the radius (one half of the length $\ell_c$) of the critical gap. The difference
between the action (\ref{a0}) on this configuration and on the trivial one is
exactly the expression for the exponential power in Eq.(\ref{g0}), and the
condition for applicability of the effective action (\ref{a0}) requires
\be
M_0 \, R _ s = {M_0 \, \mu \over  \eps_1-\eps_2} \gg 1~.
\label{rineq}
\ee
Generally one also has $\mu \gsim M_0$, and for the strings in weakly coupled
theories $\mu \gg M_0$, so that the power in the exponent in Eq.(\ref{g0}) is
large, which justifies a semiclassical treatment.

The probability of the transition is determined\cite{Coleman,Callan,Stone} by
(the imaginary part of) the ratio of the path integrals ${\cal Z}_{12}$ and
${\cal Z}_1$ calculated with the action (\ref{a0}) around respectively the
bounce configuration and around the initial flat string:
\be
\gamma_s= {1 \over A } \, \mathrm{Im}\f{{\cal
Z}_{12}}{{\cal Z}_{1  }}~.
\label{rz}
\ee
It can also be reminded that, as explained in great detail in Ref.\cite{Callan},
that the imaginary part of ${\cal Z}_{12}$ arises from one negative mode at the
bounce configuration, and that due to two translational zero modes the
numerator in Eq.(\ref{rz}) is proportional to the total space time area $ A $
in the $(t,x)$ plane occupied by the string, so that the finite quantity is the
transition probability per unit time (the rate) and per unit length of the
string.

In order to evaluate the relevant path integrals with pre-exponential
accuracy we use the cylindrical coordinates, with $r$ and $\theta$ being the
polar variables in the $(t,x)$ plane (of the bounce), and $z$ being the
transverse coordinate. We consider only one transverse coordinate, since the
effect of each of the extra dimensions factorizes, so that the corresponding
generalization is straightforward. We further assume, for definiteness, that the
space-time boundary in the $(t,x)$ plane is a circle of large radius $L$, where
the
boundary condition for the string is $z(r=L)=0$. The small deviations of the
string configuration from the bounce, illustrated in Fig.3, can be parametrized
by the radial ($f$) and transverse ($\x$) shifts of the boundary between the
string phases:
\be
r(\theta) = R _ s + f(\theta)\,,~~~~z(\theta)=\x(\theta)~,
\label{bv}
\ee
and by the variations of the surfaces of the two string phases: $z_1(r,\theta)$ and
$z_2(r,\theta)$, where, naturally,
\be
z_1(R _ s,\theta)= z_2(R _ s,\theta)=\x(\theta)~.
\label{bc}
\ee

\begin{figure}[ht]
  \begin{center}
    \leavevmode
    \epsfxsize=12cm
    \epsfbox{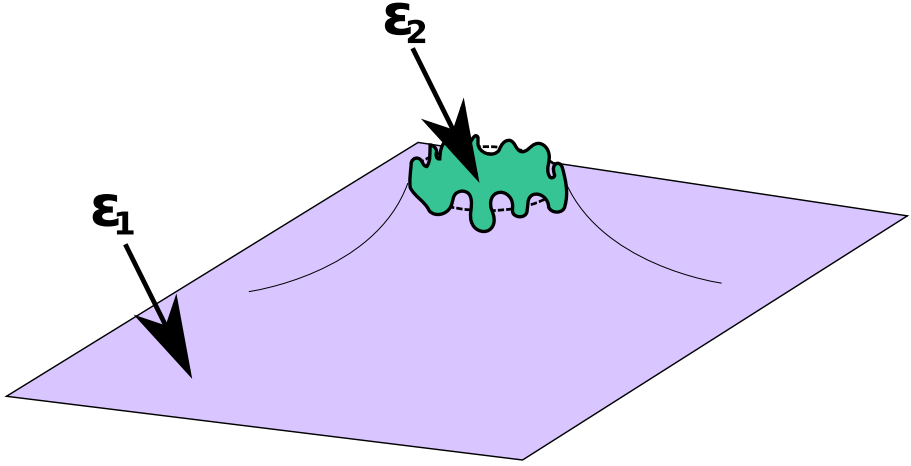}
    \caption{Fluctuations around the bounce configuration.}
  \end{center}
\end{figure}

In terms of these variables the action (\ref{a0}) can be written in the
quadratic approximation in the deviations from the bounce as
\bea
&&S_{12}=\eps_1 \, \pi \, L^2 + {\pi \, \mu^2 \over \eps_1- \eps_2} +
{\eps_1-\eps_2 \over 2} \, \int d\theta \, \left (\dot{\x}^2+\dot{f}^2-f^2
\right )+
\nonumber \\
&& {\eps_1 \over 2} \, \int_{R _ {s} } ^ {L} r drd\theta \,\left ( z_1'\,^2+ {\dot{z_1}^2
\over r^2} \right ) + {\eps_2 \over 2} \, \int _  0 ^ { R _ s } r drd\theta \,\left (
z_2'\,^2+ {\dot{z_2}^2 \over r^2} \right )~,
\label{s12}
\eea
where the primed and dotted symbols stand for the derivatives with respect to
$r$ and $\theta$ correspondingly.

Finally, the action around a flat initial string configuration in the quadratic
approximation takes the form
\be
S_1 = \eps_1 \, \pi L^2 +{\eps_1 \over 2} \, \int_0^L r drd\theta \,\left (
z'\,^2+ {\dot{z}^2 \over r^2} \right )~,
\label{s1}
\ee
with $z(r,\theta)$ parametrizing small deviations of the string in the
transverse direction.

\subsubsection{Separating variables in the path integrals}
One can readily see that in the quadratic part of the action (\ref{s12}) the
`longitudinal' variation of the bounce boundary in the $(t,x)$ plane, described
by the function $f(\theta)$ completely decouples from the rest of the variables.
This implies that the path integral over $f$ can be considered independently of
the integration over other variables and that it enters as a factor in ${\cal
Z}_{12}$. On the other hand it is this integral that provides the imaginary part
to the partition function, and it is also proportional to the total space-time
area $ A $. Moreover, this path integral is identical to the one entering the
problem of false vacuum decay in (1+1) dimensions and we can directly apply the
result of that calculation\cite{mv85}:
\be
{1 \over  A } \, \mathrm{Im} \int {\cal D} f \, \exp \left [ - {\eps_1-\eps_2
\over 2} \, \int_0^{2\pi} \, \left ( \dot{f}^2 - f^2 \right ) \, d\theta \right]
=  {\eps_1-\eps_2 \over 2 \pi}~.
\label{dim2r}
\ee
The expression for the transition rate thus can be written in the form
\be
{d \Gamma \over d \ell} = {\eps_1-\eps_2 \over 2 \pi} \, \exp \left ( -{\pi
\,\mu^2 \over \eps_1-\eps_2} \right ) \, \, {{\tilde {\cal Z}}_{12} \over {\cal
Z}_1} ~,
\label{g2}
\ee
with the path integral ${\tilde {\cal Z}}_{12}$ running only over the transverse
variables $\x$, $z_1$ and $z_2$,
\be
{\tilde {\cal Z}}_{12} = \int {\cal D}\x \, {\cal D}z_1 \, {\cal D} z_2 \,
\exp \left (- {\tilde S}_{12} \right)
\label{tz12}
\ee
and involving only the quadratic in these variables part of the action
(\ref{s12}),
\be
{\tilde S}_{12} = {\eps_1-\eps_2 \over 2} \, \int d\theta \, \dot{\x}^2 +
{\eps_1 \over 2} \, \int _ { R _ s } ^ L r drd\theta \,\left ( z_1'\,^2+ {\dot{z_1}^2 \over
r^2} \right ) + {\eps_2 \over 2} \, \int _ 0 ^ { R _ s }  r drd\theta \,\left ( z_2'\,^2+
{\dot{z_2}^2 \over r^2} \right )~.
\label{ts12}
\ee
In the same quadratic approximation the flat string partition function ${\cal
Z}_1$ is given by
\be
{\cal Z}_1= \int {\cal D} z \, \exp \left (- S_1 \right )
\label{z1}
\ee
with $S_1$ given by Eq.(\ref{s1}) and the integral running over all the
functions vanishing at the space-time boundary: $z(L, \theta)=0$.

At this point there still is a coupling in the path integral (\ref{tz12})
between the bulk variables $z_1$, $z_2$ and the boundary variable $\x$
arising from the boundary conditions (\ref{bc}). This however is a simple issue
which is resolved by a straightforward shift of the integration variables $z_1$
and $z_2$. Namely, we write
\be
z_1(r, \theta)=z_{1c}(r, \theta) + z_{1q}(r, \theta)~,~~~~z_2(r,
\theta)=z_{2c}(r, \theta) + z_{2q}(r, \theta)~,
\label{zcq}
\ee
where $z_{1q}$ and $z_{2q}$ are the new integration variables in ${\tilde {\cal
Z}}_{12}$ and these functions satisfy zero boundary conditions,
\be
z_{1q}(R _ s, \theta)=z_{2q}(R _ s, \theta)=z_{1q}(L, \theta)=z_{2q}(L, \theta)=0~,
\label{zqbc}
\ee
while $z_{1c}$ and $z_{2c}$ are fixed (for a fixed $\x(\theta)$) functions
satisfying the boundary conditions
\be
z_{1c}(R _ s, \theta)=z_{2c}(R _ s, \theta)=\x(\theta)~, z_{1c}(L, \theta)=0~,
\label{zcbc}
\ee
which are also harmonic, i.e. satisfying the Laplace equation $\Delta z=0$.

After the specified shift of the variables we arrive at the expression for the
action (\ref{ts12}) in which the bulk and the boundary degrees of freedom are
fully separated:
\bea
{\tilde S}_{12} &=& {\eps_1-\eps_2 \over 2} \, \int d\theta \, \dot{\x}^2 + R _ s
\, \int d \theta  \,  \left ( {\eps_2 \over 2} {\partial_r} z_{2c} \left .
\right |_{r = R _ s} - {\eps_1 \over 2} {\partial_r} z_{1c} \left . \right |_{r = R _ s}
\right ) \, \x
\nonumber \\
&-& {\eps_1 \over 2} \, \int d^2r \, z_{1q} \Delta z_{1q} - {\eps_2 \over 2} \,
\int d^2r \, z_{2q} \Delta z_{2q}~.
\label{tss}
\eea
Clearly, the boundary terms in the first line here, arising from the integration
by parts, depend only on the transverse shift of the boundary $\x(\theta)$.
The partition function ${\tilde {\cal Z}}_{12}$ can thus be written as a product
of the `boundary' and the `bulk' terms:
\be
{\tilde {\cal Z}}_{12} = {\cal Z}_{12 (\rm boundary)} \, {\cal Z}_{12 (\rm
bulk)}~,
\label{z12bb}
\ee
with the ${\cal Z}_{12 (\rm bulk)}$ being given by the path integral over the
bulk variables $z_{1q}$ and $z_{2q}$ only :
\be
{\cal Z}_{12 (\rm bulk)}=\int {\cal D} z_{1q} \,  {\cal D} z_{q} \, \exp \left (
{\eps_1 \over 2} \, \int d^2r \, z_{1q} \Delta z_{1q} + {\eps_2 \over 2} \, \int
d^2r \, z_{2q} \Delta z_{2q} \right )~,
\label{z12bk}
\ee
and the boundary term
\be
{\cal Z}_{12 (\rm boundary)}=\int {\cal D}\x \, \exp \left [ - {\eps_1-\eps_2
\over 2} \, \int d\theta \, \dot{\x}^2 - R _ s \, \int d \theta  \,  \left (
{\eps_2 \over 2} {\partial_r} z_{2c} \left . \right |_{r=R _ s} - {\eps_1 \over 2}
{\partial_r} z_{1c} \left . \right |_{r=R _ s} \right ) \, \x \right]
\label{z12by}
\ee
involving integration over only the boundary values.

The subsequent calculation of the ratio of the partition functions in
Eq.(\ref{g2}) can in fact be reduced to a calculation of ${\cal Z}_{12 (\rm
boundary)}$ only. In order to achieve this reduction one should organize the
partition function ${\cal Z}_1$ in the denominator of Eq.(\ref{g2}) in a form
similar to Eq.(\ref{z12bb}) as follows. Although the flat string configuration
makes no reference to a circle of the radius $R$, the partition function
${\cal Z}_1$ can be calculated by first fixing the transverse variable $z$ at
$r=R _ s$: $z(R _ s, \theta)= \x(\theta)$ and separating the integration over the
bulk variables. Then the flat string partition function factorizes in the form
similar to Eq.(\ref{z12bb}):
\be
{\tilde {\cal Z}}_{1} = {\cal Z}_{1 (\rm boundary)} \, {\cal Z}_{1 (\rm bulk)}~,
\label{z1bb}
\ee
with ${\cal Z}_{1 (\rm bulk)}$ given a similar path integral as in
Eq.(\ref{z12bk}),
\be
{\cal Z}_{1 (\rm bulk)}=\int {\cal D} z_{1q} \,  {\cal D} z_{q} \, \exp \left (
{\eps_1 \over 2} \, \int d^2r \, z_{1q} \Delta z_{1q} + {\eps_1 \over 2} \, \int
d^2r \, z_{2q} \Delta z_{2q} \right )~,
\label{z1bk}
\ee
where, as in Eq.(\ref{z12bk}),  $z_{1q}$ and $z_{2q}$ are respectively the outer
(i.e. at $r > R _ s$) and the inner ($r < R _ s$) transverse fluctuations with zero
boundary conditions. The difference in the coefficient in the expressions
(\ref{z12bk}) and (\ref{z1bk}) for the contribution of the inner part, $\eps_2$
vs. $\eps_1$, is not essential, since the overall coefficient of the quadratic
part of the action is absorbed in the measure of integration, as can be seen by
rescaling to the corresponding canonically normalized variables $\phi =
\sqrt{\eps} \, z$.

One therefore finds that ${\cal Z}_{1 (\rm bulk)} = {\cal Z}_{12 (\rm bulk)}$,
and the ratio of the partition functions in Eq.(\ref{g2}) is in fact determined
by the ratio of the boundary terms.

\subsubsection{Regularization}
The boundary factor ${\cal Z}_{1 (\rm boundary)}$ for the flat string is
somewhat different from the one given by Eq.(\ref{z12by}) and reads as
\be
{\cal Z}_{1 (\rm boundary)}=\int {\cal D}\x \, \exp \left [ -R _ s \,{\eps_1
\over 2} \, \int d \theta \,  \left ({\partial_r} z_{2c} \left . \right |_{r=R _ s}
-  {\partial_r} z_{1c} \left . \right |_{r=R _ s} \right ) \, \x \right]~,
\label{z1by}
\ee
where the functions $z_{1c}$ and $z_{2c}$ are defined in the same way as in
Eq.(\ref{z12by}).

The latter functions can be readily found by expanding the boundary function
$\x(\theta)$ in angular harmonics:
\be
\x(\theta)={a_0 \over \sqrt{2\pi}} + {1 \over \sqrt{\pi}}\sum_{n=1}^\infty
\left [ a_n \, \cos( n \theta)+b_n \, \sin(n \theta) \right ]
\label{ft}
\ee
with $a_n$ and $b_n$ being the amplitudes of the harmonics. Then at $n \neq 0$
the harmonics of the discussed functions are found as
\be
z_{1c}^{(n)}(r, \theta)= {1 \over \sqrt{\pi}} \,\left [ a_n \, \cos( n
\theta)+b_n \, \sin(n \theta) \right ] \, {R _ s^n \over r^n}\,, ~~z_{2c}^{(n)}(r,
\theta)= {1 \over \sqrt{\pi}} \, \left [ a_n \, \cos( n \theta)+b_n \, \sin(n
\theta) \right ] \, {r^n \over R _ s^n}~,
\label{hn}
\ee
and for n=0 these are given by
\be
z_{1c}^{(0)}(r, \theta) = a_0 \, {\ln (r /L) \over \ln (R_s/L)}~, ~~~~
z_{2c}^{(0)}(r, \theta) =a_0~.
\label{h0}
\ee
Substituting these expressions for the harmonics in the equations (\ref{z12by})
and (\ref{z1by}) and performing the Gaussian integration over the amplitudes
$a_n$ and $b_n$ we find the  boundary factors in the form
\be
{\cal Z}_{12 (\rm boundary)}={\cal N} \,  \sqrt{\eps_1 \, \ln {L \over R _ s}} \,
\prod_{n=1}^\infty {1 \over (\eps_1-\eps_2) \, n^2 + (\eps_1 + \eps_2) \, n}~
\label{z12by1}
\ee
and
\be
{\cal Z}_{1 (\rm boundary)}={\cal N} \,  \sqrt{\eps_1 \, \ln {L \over R _ s}} \,
\prod_{n=1}^\infty {1 \over 2 \, \eps_1 \, n}
\label{z1by1}
\ee
with ${\cal N}$ being a normalization factor.

Clearly, each of the formal expressions (\ref{z12by1}) and (\ref{z1by1})
contains a divergent product, and their ratio is also ill defined, so that our
calculation requires a regularization procedure that would cut off the
contribution of harmonics with large $n$. A regularization of high harmonics is
also required on general grounds. Indeed, as previously mentioned, our
consideration using the effective string action (\ref{a0}) is only valid for
smooth deformations of the string, i.e. as long as the relevant momenta are
smaller than the mass scale $M_0$ for excitation of the internal degrees of
freedom within the thickness of the string. For an $n$-th harmonic the relevant
momentum is $k \sim n/R _ s$ so that the applicability of the effective low energy
action requires a cutoff at $n \ll M_0R _ s$.   In order to perform such
regularization we use the standard Pauli-Villars method and introduce a
regulator field $Z$ with negative norm and the action corresponding to the
quadratic part of the Nambu-Goto expression (\ref{a0}) for small $z$:
\be
S_R=\f{\eps_1 - \eps_2}{2} \int d\theta \dot{\x_R}^2+  {\eps_1 \over 2} \,
\int_{A_1} d^2r \left [ (\partial_\mu Z)^2 + M^2 \, Z^2
\right ] + {\eps_2 \over 2} \, \int_{A_2} d^2r \left [ (\partial_\mu Z)^2 + M^2
\, Z^2 \right ]
\label{sr}
\ee
with $M$ being the regulator mass, which physically should be understood as
satisfying the condition $M \ll M_0$ and still being much larger than the
relevant scale in the discussed problem, in particular $MR _ s \gg 1$.

The regularized expression for the ratio of the boundary terms in ${\cal
Z}_{12}$ and ${\cal Z}_{1}$ thus takes the form
\be
{{\cal Z}_{12 (\rm boundary)} \over {\cal Z}_{1 (\rm boundary)}} \longrightarrow
{\cal R}=
\left [ {{\cal Z}_{12 (\rm boundary)} \over {\cal Z}^{(R)}_{12 (\rm boundary)}}
\right ] \, \left [ {{\cal Z}_{1 (\rm boundary)} \over {\cal Z}^{(R)}_{1 (\rm
boundary)}} \right ]^{-1}~,
\label{reg}
\ee
where we introduced the notation ${\cal R}$ for the regularized ratio, and the
regulator partition functions ${\cal Z}^{(R)}_{12 (\rm boundary)}$
and ${\cal Z}^{(R)}_{1 (\rm boundary)}$ are determined by the same expressions
as in Eqs.(\ref{z12by}) and (\ref{z1by}) with the `outer' and `inner' functions
$z_{1c}$ and $z_{2c}$ being replaced by their regulator counterparts $Z_{1c}$
and $Z_{2c}$ which still satisfy the boundary conditions similar to
(\ref{zcbc}):
\be
Z_{1c} (R _ s, \theta)= Z_{2c}(R _ s, \theta)=\x_R(\theta)~,
\label{zrbc}
\ee
but are the solutions of the Helmholtz rather than Laplace equation: $(\Delta -
M^2) Z=0$.

The solutions of the latter equation fall off exponentially at the scale
determined by $M$, and for our purposes only the behavior near the circle $r=R _ s$
is needed. For this reason we write the equation for the radial part of the
$n$-th angular harmonic $Z_n(r)$,
\be
Z_n''+{1 \over r} \, Z_n' -{n^2 \over r^2} \, Z_n - M^2 \, Z_n = 0~,
\label{zneq}
\ee
and parametrize the radial coordinate as $r=R _ s+x$, and treat the parameter
$(x/R _ s)$ as small, since the scale for the variation of the solution is $x \sim
1/\sqrt{M^2+n^2/R _ s^2}$. This approach yields an expansion of the regulator action
associated with the boundary at $r=R$ in powers of $1/\sqrt{(MR_s)^2 + n^2}$. With
the accuracy required in the present calculation, the (normalized to one at
$r=R _ s$) solution to Eq.(\ref{zneq}) is found in the first order of expansion in
$(x/R _ s)$ as
\be
Z_n(R _ s+x)=\left ( 1- {1 \over 2} \, {(M R _ s)^2 \over  (MR _ s)^2+n^2} \, {x \over R _ s}
\right ) \, \exp \left ( -\sqrt{(MR _ s)^2+n^2}
 \, {|x| \over R _ s} \right )~.
\label{znsol}
\ee

Using the form of the solutions for the harmonics of the regulator field given
by Eq.(\ref{znsol}) and the expressions (\ref{z12by1}) and (\ref{z1by1}), one
can
write the regularized ratio of the boundary partition functions (\ref{reg}) as
\bea
{\cal R} &=& \sqrt{\eps_1+\eps_2 \over 2 \eps_1} \, \left [ \prod_{n=1}^\infty
{n^2 +b \, \sqrt{(MR _ s)^2+n^2} \over n^2 + b \, n} \right ] \, \left [
\prod_{n=1}^\infty {n \over \sqrt{(MR _ s)^2+n^2}} \right ]\, \times \nonumber \\
&&\prod_{n=1}^\infty \left \{ 1 +  {1 \over 2} \, {(MR _ s)^2 \over \left
[(MR _ s)^2+n^2 \right ] \, \left [ n^2 +b \, \sqrt{(MR _ s)^2+n^2}\, \right ] } \right
\}~,
\label{crprod}
\eea
where we introduced the notation
\be
b={\eps_1+\eps_2 \over \eps_1-\eps_2}~,
\label{bdef}
\ee
and the last product factor in Eq.(\ref{crprod}) arises from the first term of
expansion in $(x/R)$ in the pre-exponential factor in Eq.(\ref{znsol})

\subsubsection{Calculating the products}
Each of the products in Eq.(\ref{crprod}) is finite at a finite $M$ and can be
calculated separately. We start with the most straightforward one, which is
directly given by an Euler's formula:
\be
\prod_{n=1}^\infty {n \over \sqrt{(MR _ s)^2+n^2}} = \sqrt{\pi M R _ s \over \sinh (\pi
MR _ s)} \longrightarrow \sqrt{2 \pi M R _ s } \, \exp \left ( - {\pi M R _ s \over 2}
\right )~,
\label{mprod}
\ee
where the last transition corresponds to the limit $M R _ s \gg 1$.

The other two factors in Eq.(\ref{crprod}), the first and the last, generally
depend on the relation between the parameters $b$ and $MR _ s$, or equivalently
between $(\eps_1+\eps_2)$ and $\mu M$. We find however that the latter product
is equal to one at $MR _ s \gg 1$ independently of $b$. In particular in the
nontrivial case of $b \ll MR _ s$ we find
\bea
&&\ln  \prod_{n=1}^\infty \left . \left \{ 1 +  {1 \over 2} \, {(MR _ s)^2 \over
\left [(MR _ s)^2+n^2 \right ] \, \left [ n^2 +b \, \sqrt{(MR _ s)^2+n^2}\, \right ] }
\right \} \right|_{MR _ s \to \infty} \to
\nonumber \\
&&\left . \left \{ {(MR _ s)^2 \over 2} \, \int_{n_0}^\infty dn \, \left [(MR _ s)^2+n^2
\right ]^{-1} \,
 \left [ n^2 +b \, \sqrt{(MR _ s)^2+n^2}\, \right ]^{-1} \right \}  \right|_{MR _ s \to
\infty} \to 0~,
\label{f1}
\eea
where the lower limit in the integral is any number $n_0$ that is finite in the
limit $MR _ s \to \infty$.

The dependence of the first product factor in Eq.(\ref{crprod}) on the ratio
$(\eps_1+\eps_2)/(\mu M)= b/(MR _ s)$ is essential and we consider two limiting
cases when this ratio is much bigger than one and when it is very small. In the
former case, i.e. for $b \gg MR _ s$, the first product in Eq.(\ref{crprod}) becomes
reciprocal of the second, and one finds
\be
{\cal R}|_{b \gg MR _ s \gg 1}=1~.
\label{rl}
\ee
(Clearly one can also safely make the replacement $(\eps_1+\eps_2)/(2 \eps_1)
\to 1$ at $b \gg 1$.)

The behavior of ${\cal R}$ in the opposite limit, i.e. at $b \ll MR_s$, turns out
to be significantly more interesting. Using the Euler-Maclaurin summation
formula for the logarithm of the first product in Eq.(\ref{crprod}) we find in
the limit $MR _ s \gg 1$ and $MR _ s \gg b$:
\be
\prod_{n=1}^\infty {n^2 +b \, \sqrt{(MR _ s)^2+n^2} \over n^2 + b \, n} =
{\Gamma(b+1) \over 2 \pi \, \sqrt{b MR _ s}} \, \exp \left [ \pi \, \sqrt{b MR _ s} - b
\, \ln (MR _ s)- b \, (1-\ln 2) \right ]~.
\label{r1}
\ee
Being combined with the expression (\ref{f1}) this yields the formula
\be
{\cal R}=  \sqrt {\eps_1+\eps_2 \over 2 \eps_1} \, {\Gamma(b+1) \over \sqrt{2
\pi \, b}}\,  \exp \left [- {\pi \over 2} \, MR_s + \pi \, \sqrt{b MR_s} - b \, \ln
(MR_s)- b \, (1-\ln 2) \right ]~,
\label{rh}
\ee
which contains an essential dependence on the regulator mass parameter $M$. We
will show however that all such dependence in the phase transition rate can be
absorbed in renormalization of the parameter $\mu$ in the leading semiclassical
term.

\subsubsection{Renormalization of $\mu$}
The parameter $\mu$ is defined in the action (\ref{a0}) as the coefficient in
front of  the length of the  boundary between the world sheets for two phases of
the string. Generally this parameter gets renormalized by the quantum
corrections, and in order to find such renormalization at the level of first
quantum corrections, one needs to perform the path integration using the
quadratic part of the action around a configuration, in which the length of the
interface is an arbitrary parameter. For a practical calculation of this effect
we consider a Euclidean space configuration, shown in Fig.4, with the string
lying flat along the $x$ axis, and the interface between the two phases being at
$x=0$. The length of the world line for the boundary is thus the total size $T$
of the world sheet in the time direction. It should be mentioned that such
configuration with different string tension on each side of the boundary is not
stationary for the action (\ref{a0}). However it can be `stabilized' by a source
term (external force) depending on the coordinate $x(t)$ of the boundary: $\int
J(t) x(t) \, dt$, which term does not affect the quadratic in fluctuations part
of the action.

\begin{figure}[ht]
  \begin{center}
    \leavevmode
    \epsfxsize=12cm
    \epsfbox{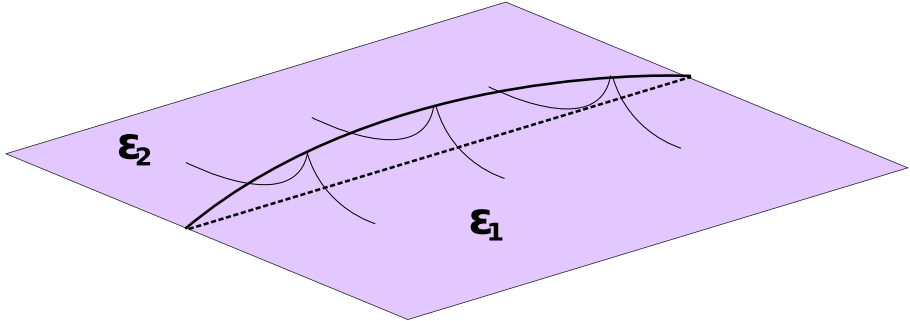}
    \caption{The configuration for the calculation of the renormalization of
$\mu$.}
  \end{center}
\end{figure}

The Gaussian path integral over the transverse coordinates $z(x,t)$ is then to
be calculated with zero boundary conditions at the edges of the total world
sheet. Using the notation $\x(t)$ for the transverse shift of the boundary,
this condition implies $\x(0)=\x(T)=0$, so that the function $\x(t)$
has the Fourier expansion of the form
\be
\x(t)=\sqrt{2 \over T} \, \sum_{n=1}^\infty a_n \, \sin \left ( {\pi \, n \,
t \over T} \right )~,
\label{zt}
\ee
and a similar expansion applies to the regulator boundary function $\x_R(t)$.
The part of the effective action associated with the boundary is determined by
the functions $z_c(x,t)$ for the transverse shift of the string and the
corresponding regulator functions $Z_c(x,t)$ that satisfy the equations
\be
\Delta z_c=0 ~~~{\rm and} ~~~(\Delta - M^2) \,Z_c =0~,
\label{laphel}
\ee
and the boundary conditions
\be
z_c(0,t)=\x(t)~,~~~~Z_c(0,t)=\x_R(t)
\label{bcf}
\ee
as well as zero boundary conditions at the edges of the total world sheet. One
can readily find these functions for each harmonic of the boundary values
$\x$ and $\x_R$, using the solutions for $z_c$ and $Z_c$
in each harmonic:
\be
z_c^{(n)}(x,t) = \exp \left ( -|x| \,   {\pi \, n \over T}\right ) \, \sin \left
( {\pi \, n \, t \over T} \right ) ~,
\label{zcf}
\ee
and
\be
Z_c^{(n)}(x,t) = \exp \left [ -|x| \, \sqrt{ \left ( {\pi \, n \over T}\right
)^2 + M^2} \, \right ] \, \sin \left ( {\pi \, n \, t \over T} \right )~.
\label{zrcf}
\ee

In order to separate the boundary effect in the path integral around the
considered configuration from the bulk effects, we again divide it by the path
integral around the configuration where the whole world sheet is occupied by the
same phase of the string. The latter phase can be chosen with either of the
tensions, or with an arbitrary tension $\eps$. Such division results, as
previously, in the cancellation of the bulk contributions, and
the remaining part of the effective action associated with the boundary is
written in
terms of the regularized path integral over the boundary function $\x$ as
\bea
&&\mu_R \, T = \mu \, T - \nonumber \\
&&\ln  {\int {\cal D} \x  \, \exp \left \{-{1 \over 2} \, \int dt \left [ \mu
\dot{\x}^2 + \x(t) \, \left ( \eps_2  \, z_c'(x,t)|_{x \to -0} -  \eps_1
\, z_c'(x,t)|_{x \to +0} \right ) \right ] \right \} \over  \int {\cal D}
\x_R  \, \exp \left \{ -{1 \over 2} \, \int dt \left [ \mu  \dot{\x_R}^2 +
\x_R(t) \, \left (\eps_2 \, Z_c'(x,t)|_{x \to -0} -  \eps_1  \, Z_c'(x,t)|_{x
\to +0} \right ) \right ] \right \} } + \nonumber \\
&&\ln  {\int {\cal D} \x  \, \exp \left \{- \eps  \, \int dt  \x(t) \,
z_c'(x,t)|_{x \to -0}  \right \} \over  \int {\cal D} \x_R  \, \exp \left \{
-  \eps \, \int dt   \x_R(t) \, Z_c'(x,t)|_{x \to -0}  \right \} }~,
\label{muet}
\eea
where $\mu_R=\mu + \delta \mu$ is the renormalized mass parameter.  The
correction to $\mu$ can thus be written in the form
\be
\delta \mu= -{1 \over 2 \, T} \ln \left \{ \prod_{n=1}^\infty {n^2 +{\tilde b}
\, \sqrt{(MT/\pi)^2+n^2} \over n^2 + {\tilde b} \, n} \, \prod_{n=1}^\infty {n
\over \sqrt{(MT/\pi)^2+n^2}}\right \}~,
\label{ptb}
\ee
where
\be
{\tilde b}= { \eps_1+\eps_2 \over \mu} \, {T \over \pi}~.
\label{tbd}
\ee
In the limit $\eps_1+\eps_2 \ll \mu M$ one can directly apply the results in
Eqs.(\ref{mprod}) and (\ref{r1}) for evaluation of the expression (\ref{ptb})
and write
\bea
\delta \mu &=& -{1 \over 2 \, T} \left ( {\tilde b} \ln {\tilde b} - {\tilde b}
- {M T \over 2} + \pi \sqrt{ {\tilde b} \, {M T \over \pi}} - {\tilde b} \ln {MT
\over \pi} - {\tilde b} \, (1- \ln2) \right )\nonumber \\
&=&{M \over 4} - {1 \over 2} \, \sqrt{(\eps_1+ \eps_2) \, M \over \mu} -
{\eps_1+\eps_2 \over 2 \pi \mu} \, \ln {2 \, (\eps_1+ \eps_2) \over \mu \, M}~,
\label{dmu}
\eea
where a use is made of the Stirling formula
$$\ln {\Gamma({\tilde b}+1) \over \sqrt{2 \pi {\tilde b}}} \to {\tilde b} \,
(\ln {\tilde b} - 1)~,$$
considering that ${\tilde b}$ is proportional to large $T$.

In the limiting case where $\eps_1+\eps_2 \gg \mu M$ the correction $\delta \mu$
vanishes, so that the renormalization effect is negligible.

\subsubsection{The result}
We can now assemble all the relevant elements into a formula for the rate of
the considered transition. Clearly, the path integration over the  variables $z$
factorizes for each of the $(d-2)$ transverse dimensions, so that the expression
for the decay rate takes the form
\be
\gamma_s ={\eps_1-\eps_2 \over 2 \pi} \,{\cal R}^{d-2} \, \exp
\left ( -
\, {\pi \, \mu^2 \over \eps_1-\eps_2} \right )~,
\label{gfr}
\ee
where $\mu$ is the zeroth order mass parameter. In the case of large string
tension, $\eps_1 + \eps_2 \gg \mu \, M_0$, the `bare' $\mu$ coincides with the
renormalized one, and the factor ${\cal R}$ is equal to one. It can be noted
that
the resulting obvious expression for the rate is also correctly given by
Eq.(\ref{gf}) as soon as the factor $F$ in Eq.(\ref{ff}) is taken in the limit
$\eps_1-\eps_2 \ll \eps_1+\eps_2$: $F \to 1$, which limit, as previously
discussed, is mandated in this case.

In the opposite limit of heavy $\mu$, $\eps_1 + \eps_2 \ll \mu \, M_0$, both the
expression (\ref{rh}) depends on the regulator mass $M$ and the $M$-dependent
renormalization of $\mu$ is essential. Taking into account that each of the
transverse dimensions contributes additively to $\delta \mu$ and expressing in
Eq.(\ref{gfr}) the bare $\mu$ through the renormalized one: $\mu=\m_R-\delta
\mu$, one readily finds that the dependence on the regulator mass M cancels in
the transition rate, and one arrives at the formula given by Eq.(\ref{gf}) and
Eq.(\ref{ff}).

The formula (\ref{ff}) is applicable for arbitrary ratio of the string tensions
$\eps_2/\eps_1$. In particular it can be also applied at $\eps_2 = 0$, in which
case the considered transition describes a complete breaking of the string.

It can be also noted that numerically the factor $F$ depends very moderately on
the ratio of the tensions and changes approximately linearly between $F(0) =
e/\sqrt{4 \pi} = 0.7668\ldots$ and $F(1)=1$. We thus conclude that the
two-dimensional expression ${(\eps_1-\eps_2) / (2 \pi)}$ for the pre-exponential factor in the
transition rate provides a fairly accurate approximation in higher dimensions as well, as long as the exponential factor is expressed in terms of the physical renormalized mass $\mu_R$.

There is however an interesting methodical point pertaining to the considered
here problem. Indeed, as was already mentioned, the difference from the problem
of particle creation by external electric field is in that the motion of the
ends of the string involves in addition to the mass $\mu$ also an adjacent part
of the string. In terms of the calculation of the path integrals around the
bounce the difference is in that the spectrum of soft modes in the particle
creation problem (as well as in that of the two-dimensional false vacuum decay)
consists only of the modes associated with one-dimensional world line of the
boundary of the bounce. The entire pre-exponential factor can then be found
using the effective low energy action for these modes\cite{mv85}. In the
considered here string transition there are also low modes in the bulk of the
world sheet of the string, and there is no parametric separation of their
eigenvalues from those of the modes associated with fluctuations of the
boundary. In the presented calculation the separation of the boundary and bulk
variables is achieved through an `artificial' organization of the normalization
partition function for a flat string into boundary and bulk factors ${\cal
Z}_{\rm boundary}$ and ${\cal Z}_{\rm bulk}$. The bulk contribution then cancels
in the ratio of the partition functions near the bounce and near a flat string,
so that the remaining calculation is reduced to considering the integrals over
the boundary functions only. One can also readily notice that the additional
contribution to the action from the boundary terms as e.g. those with the
functions $z_{1c}$ and $z_{2c}$  in Eq.(\ref{z12by}) corresponds to precisely
the effect of `dragging' of the string by its end.

%\newpage
\subsection{String transition at non-zero temperature \label{sect2}} 

The formula in Eq.(\ref{rz}) corresponds to a calculation of the decay rate as
the imaginary part of the energy of the initial string. At a finite temperature
$T$ the corresponding relevant quantity is the imaginary part of the free
energy\cite{Langer}, which one can calculate in the Euclidean space by
considering periodic in time configurations with the period $\beta = T^{-1}$. In
other words the thermal calculation corresponds to the path integration in the
Euclidean space-time having the topology of a cylinder. The nucleation rate is
then described by the same formula (\ref{rz}) with the action and the area being
calculated over one period, $ A = X \, \beta$, where $X$ is the length of the string 
along the spatial dimension.

We consider only sufficiently small temperatures
$\b>\ell_c$, at which temperatures we show the thermal effects behaving as
powers of $T$, which distinguish the string process from the decay of metastable
vacuum\cite{Garriga94}. We also treat the length $X$ of the metastable string as
the largest length parameter in the problem, so that $\b \ll X$. Under these
conditions the bounce corresponding to the action (\ref{a0}) is the same as at
zero temperature, except that it is placed on a long cylinder rather than on a large
flat plane (Fig.~5).
\begin{figure}[ht]
  \begin{center}
    \leavevmode
    \epsfxsize=7cm
    \epsfbox{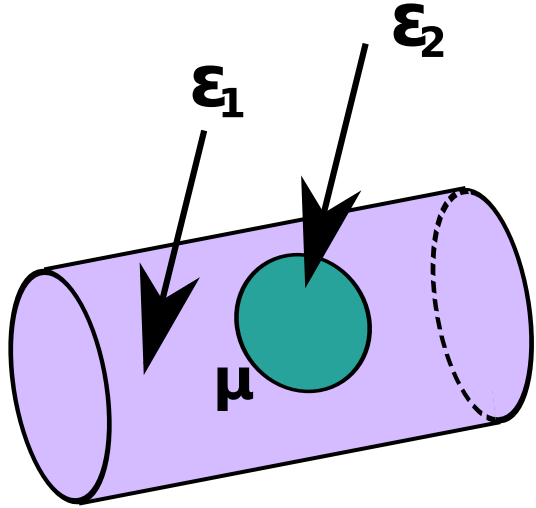}
    \caption{Bounce configuration for nonzero temperature}
  \end{center}
\label{bounce}
\end{figure}

As above we aim at calculating the path integral over the variations of the string around
the bounce configuration as illustrated in Fig.~6. The coordinates on the
cylinder (or on the plane where all the points separated by $n\b$ along
Euclidean time are identified) are $t$, $x$, with $t$ being the periodic time
coordinate, and the coordinate orthogonal to the surface of the cylinder is $z$.
The boundary conditions for the configurations over which we integrate are
\be
z \left ( t,x= \pm {X \over 2} \right )=0, \,~~~~~z(t+\b,x)=z(t,x)~.
\label{boundary_conditions}
\ee
In polar coordinates $(r,\theta)$ in the $(t,x)$-plane one 
finds the variation of the bounce action (\ref{a0}) similar to (\ref{s12}) while for the flat string one has the expression analogous to that in Eq.(\ref{s1}).
\begin{figure}[ht]
  \begin{center}
    \leavevmode
    \epsfxsize=7cm
    \epsfbox{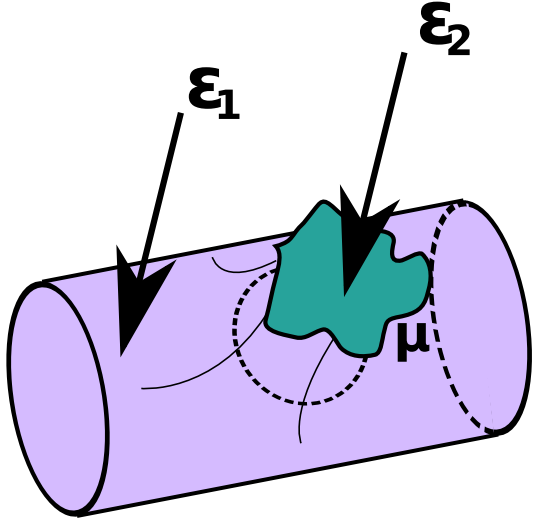}
    \caption{Variation of the bounce configuration}
  \end{center}
\label{variation}
\end{figure}

As it was shown for the case of zero temperature the partition functions factorize, therefore the expression for a decay rate is given by (\ref{g2}), with the only difference for a nonzero temperature case that the boundary conditions for outer solution are 
\be
z_{1c}(t+\b,x)=z_{1c}(t,x),
\ee
and
\be
z_{1c} \left (t,x=\pm {X \over 2} \right )=0
\label{bcz1}
\ee
while the inner solution $z_{2c}$ is required to be regular inside the disk.

In order to do the path integrals as before one can expand the boundary function
$\x(\theta)$ in angular harmonics. For the inner solution $z_{2c}$ to the Laplace equation, i.e. at $r \le R _ s$ one
finds no difficulty in finding the harmonics matching the boundary function at
the interface (\ref{hn}). For the outer solution $z_{1c}$ however there is a difficulty due to the mismatch between the symmetry of the boundary and of the
periodicity conditions. It is impossible to choose the solution to the Laplace
equation for outer string bulk variable to be $z_{1c}^{(n)}(r,\t)\sim r^{-n}$,
since it is not periodic in time. In this situation in order to have a periodic
solution one can perform a periodic mapping of the cylinder on the plane and
consider the outer solution of the Laplace equation as the sum of the solutions
produced by a ``source" at each period as illustrated in Fig.~7. Introducing the
complex variable $w=t+i\,x$, we construct the solutions for the functions
$z_{1c}(r,\theta)$ using the harmonic real and imaginary parts of the following
basis set of periodic functions, satisfying the boundary condition (\ref{bcz1})
at large $|x|$,
\bea
g_0(w)&=&\ln\l[\sin\l(\f{\pi\,w}{\b}\r)\r] - { \pi \, X \over 2 \b} - i \, {\pi
x \over X} + \ln 2~, \nonumber \\
g_1(w)&=&\f{\pi\,R _ s}{\b}\,\cot\l(\f{\pi\,w}{\b}\r) +i \, {2\, \pi \, R _ s \,x \over
\b \,X}~,\nonumber \\
{g}_k(w)&=&\f{R _ s^k}{w^k}+\sum_{n=1}^\infty\l[\f{R _ s^k}{(w-n\b)^k}+\f{R _ s^k}{(w+n\b)^k
}\r],~~~\textrm{for $k>1$}~.
\label{syst}
\eea
\begin{figure}[ht]
  \begin{center}
    \leavevmode
    \epsfxsize=10cm
    \epsfbox{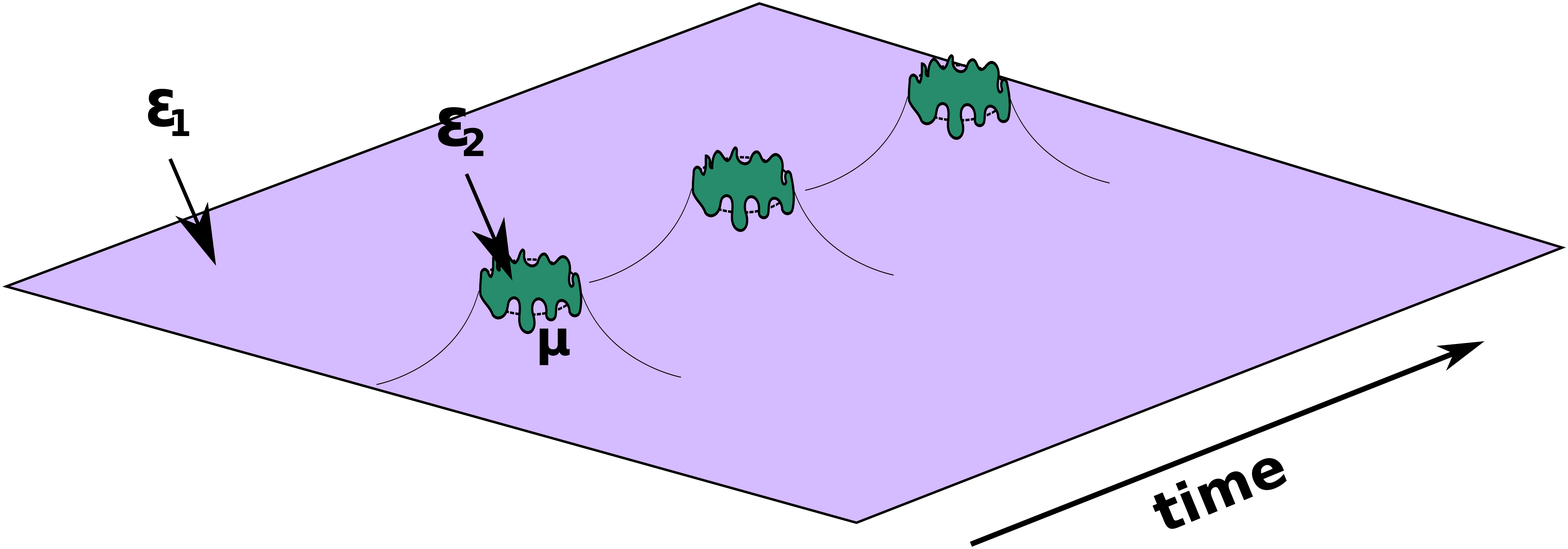}
    \epsfxsize=6cm
    \epsfbox{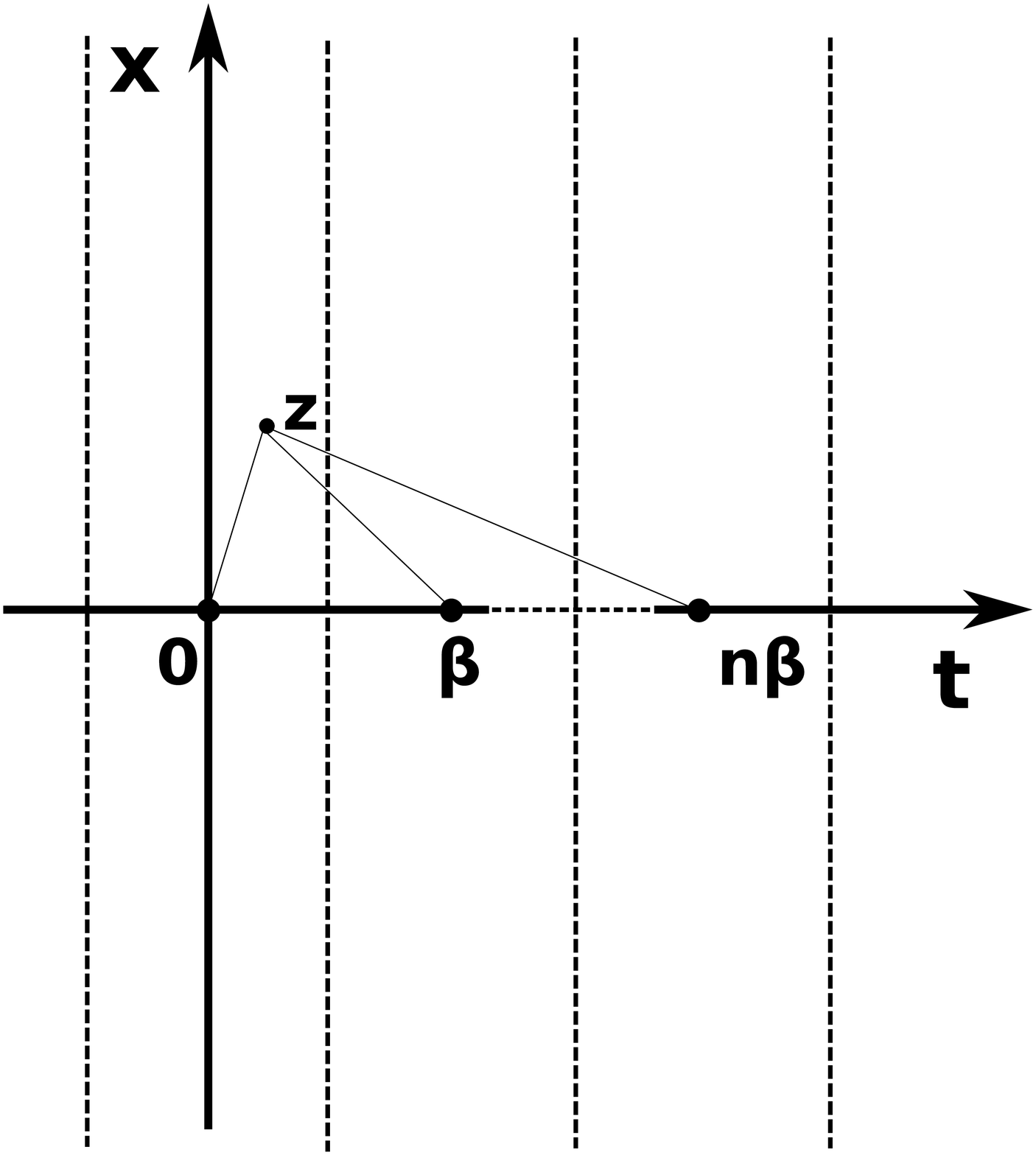}
    \caption{Periodic configuration on the plane}
  \end{center}
\label{periodic_plane}
\end{figure}

Clearly, these functions are periodic in the Euclidean time by construction, with
the period $\b$. Also the functions $g_k(w)$ with $k > 1$ are analytic complex
functions, so that their real and imaginary parts are harmonic. In the functions
$g_0(w)$ and $g_1(w)$ the explicit dependence on $x$, introducing
non-analyticity, is linear and is thus also harmonic, so that their real and
imaginary parts do satisfy the Laplace equation.

An arbitrary periodic outer solution to the Laplace equation, satisfying the
boundary condition (\ref{bcz1}) at large $|x|$ can be expanded in the series
\be
z_{1c}(r,\t)=A_0\mathrm{Re}\,g_0(w)+\sum_{k=1}^\infty
A_k\mathrm{Re}\,g_k(w)+B_k\mathrm{Im}\,g_k(w),
\ee
The disadvantage of this set of solutions (hence of such an expansion) is that
it is not orthogonal, so that the expression for the action up to quadratic
terms is not diagonal in this basis. Therefore, the calculation of the integral
is not just a calculation of the product of eigenvalues. To find the integral
over the amplitudes of the Fourier harmonics one has to express the amplitudes
of the solutions chosen $A_k$,$B_k$ in terms of the amplitudes $a_k$, $b_k$. The
relation between the coefficients $A_k$,$B_k$ and $a_k$, $b_k$ can be found from
the matching condition (first relation in \ref{zcbc}) on the interface
\be
{a_0 \over \sqrt{2\pi}} + {1 \over \sqrt{\pi}}\sum_{n=1}^\infty
\left [ a_n \, \cos( n \theta)+b_n \, \sin(n \theta) \right
]=A_0\,g_0(w)\Big|_{r=R_s}+\sum_{k=1}^\infty\l[A_k\mathrm{Re}g_k(w)+B_k\mathrm{Im}
g_k(w)\r]\Big|_{r=R_s}.
\ee
One can readily notice that the function $g_0$ contains a large constant term,
proportional to $X$, which totally dominates the matching condition for the
$a_0$ mode at $r=R _ s$, so that
\be
{a_0 \over \sqrt{2\pi}}=-A_0\,\f{\pi\,X}{2 \b}\,\l[1+O(R/X)\r].
\label{a0A0}
\ee
For this reason the effect of the mixing between $a_0$ and higher modes is
suppressed by inverse powers of $X$ and can be ignored in the limit of a long
string. For this reason in considering the mixing of the modes in the following
calculation we keep only $k\neq0$. Furthermore the linear in $x$ terms in the
functions $g_0$ and $g_1$ are suppressed at $r=R _ s$ by the factor $R _ s/X$ and can also be 
disregarded.

In what follows we consider the expansion of the functions $g_k$ at $r=R _ s$ in
powers of $R _ s/\b$, which expansion, as will be seen later, converges at $R _ s <
\b/2$.
Using
\be
(1+x)^{-k}=1+\sum_{p=1}^{\infty}(-1)^px^pC_{p+k-1}^p,
\ee
where $C_{p+k-1}^p$ are the binomial coefficients,
\be
C_{p+k-1}^p=\f{(p+k-1)!}{p!(k-1)!}
\ee
and also a definition of the Riemann $\zeta$-function
\be
\zeta(k)=\sum_{n=1}^\infty n^{-k}~,
\ee
we find for $k\neq0$
\be
g_k(w)=\f{R _ s^k}{w^k}+\sum_{p=1}^\infty d_{pk}\l(\f{w}{R _ s}\r)^p,
\label{g_exp}
\ee
with
\be
d_{pk}=\l[(-1)^k+(-1)^p\r]\l(\f{R}{\b}\r)^{k+p}\,\zeta(p+k)\,C_{p+k-1}^p.
\label{D_matrix}
\ee
We have omitted in the expression (\ref{g_exp}) a constant term, which describes
the mixing with the $a_0$ mode as well as a term explicitly proportional to
$R _ s/X$.
Using the expansion (\ref{g_exp}) for $g_k(w)$ and considering the real part of
the functions $g_k(w)$, one can find the coefficients $a_l$ in terms of $A_k$
\be
a_l=\f{1}{\pi}\sum_kA_k\int_0^{2\pi}\mathrm{Re}g_k\cos(l\t)d\t=
A_l+\sum_{k=1}^\infty A_k\,d_{lk}~,
\ee
or in the matrix form
\be
A=(1+D)^{-1}\, a~,
\label{cos_solution}
\ee
where matrix $D$ has elements $d_{lk}$. Similarly for the imaginary part one
gets
\be
b_l=\f{1}{\pi}\sum_kB_k\int_0^{2\pi}\mathrm{Im}g_k\cos(l\t)d\t=B_l-\sum_kB_kd_{l
k}
\ee
and
\be
B=(1-D)^{-1} \,b~.
\ee
As usual, the contributions from the $\cos$ and $\sin$ modes are independent, and we
consider the contribution to the boundary term from the
even ($\cos$) harmonics first
\bea
&&-\int_0^{2\pi}\l.\l(R _ sg^{(R)}\p_rg^{(R)}\r)\r|_{R _ s}d\t=\int_0^{2\pi}d\t\l[\sum_la_l\
cos(l\t)\r]
\sum_k\l[k\l(\f{R _ s}{r}\r)^k\cos(k\t)-\sum_{p=1}^\infty pd_{pk}\r] = \nonumber \\
&& \sum_kA_k\l[ka_k-\sum_ppa_pd_{pk}\r]=
\sum_{k,p}a_p\l[k\delta_{pk}-pd_{pk}\r]A_k~.
\label{boundary_term_sum}
\eea
Introducing the matrix
\be
\hat{N}=\mathrm{diag}(1,2,\dots,n,\dots)~,
\label{Nmatr}
\ee
one can rewrite the expression (\ref{boundary_term_sum}) in the matrix form
\be
-\int_0^{2\pi}\l.\l(R _ sg^{(R)}\p_rg^{(R)}\r)\r| _ {R _ s}d\t=a\hat{N}(1-D)A~.
\ee
A substitution in this expression of the solution for $A$ in terms of $a$
(\ref{cos_solution}) leads to
\be
-\int_0^{2\pi}\l.\l(R _ sg^{(R)}\p_rg^{(R)}\r)\r| _ { R _ s }d\t=a\hat{N}(1-D)(1+D)^{-1}a~.
\ee
Clearly, for the odd ($\sin$) modes one gets the same expression with the
replacement $D\to-D$. Collecting all the terms together one can write the result
for the boundary partition functions (\ref{z12by}) and (\ref{z1by}) as
\bea
{\cal Z}_{12 (\rm
boundary)}&=&\mathrm{Det}\l[(\e_1-\e_2)\hat{N}^2+\e_2\hat{N}+
\e_1\hat{N}(1-D)\f{1}{1+D}\r]^{-1/2}\cdot\{D\to-D\} \nonumber \\
{\cal Z}_{1 (\rm
boundary)}&=&\mathrm{Det}\l[\e_1\hat{N}+\e_1\hat{N}(1-D)\f{1}{1+D}\r]^{-1/2}
\cdot\{D\to-D\},
\label{z12z1d}
\eea
where $\{D\to-D\}$ means that one should take the expression and make the
replacement $D\to-D$.

The zero temperature limit for the probability rate $\gamma _ s$ formally
corresponds to setting $D \to 0$. Thus one can use the result for the zero
temperature decay rate in Eq.(\ref{gfr}), and concentrate on a calculation of the
thermal catalysis factor $K$ defined as
\be
\left . \f{d\Gamma}{d\ell} \right |_{T}={\cal K}_s \, \gamma_s\,.
\ee
In a $d$-dimensional theory, i.e. with $d-2$ transverse dimensions, the catalysis factor can be written as ${\cal K}_s=G^{d-2}$, where $G$ is the factor per each transverse direction given by
\be
G=
\f{{\cal Z}_{12 (\rm
boundary)}}{{\cal Z}_{12 (\rm boundary)}^{D=0}}
\,
\f{{\cal Z}_{1 (\rm boundary)}^{D=0}}{{\cal Z}_{1 (\rm boundary)}}\,.
\ee
According to Eq.(\ref{z12z1d}) it is a matter of simple algebra to express the
factor $G$ in terms of the matrix $D$:
\be
G=\Det\l[1-\l(\f{\hat{N}-1}{\hat{N}+b}\,D\r)^2\r]^{-1/2}
\label{gen_formula}
\ee
with $b$ defined in Eq.(\ref{bdef}).

\subsubsection{Analysis of the general formula}
In this section we consider in some detail the temperature effect in the string
transition rate described by our general formula (\ref{gen_formula}) in the
situation where the inverse temperature is larger then the diameter of the
classical configuration (bounce): $\b>2R _ s$. We first notice that due to the
presence of the factor $(\hat{N}-1)$ the first elements from the first row and
the first column of the matrix $D$, $d_{1k}$ and $d_{p1}$, enter the  expression
$[(\hat{N}-1) \, D]^2$ with zero coefficients, so that the final result
(\ref{gen_formula}) does not depend on them. Furthermore, one can see from
Eq.(\ref{D_matrix}) that the matrix element $d_{pk}$ is not equal to zero only
if the indices $p$ and $k$ have the same parity. Hence, there is no mixing
between the amplitudes with even ($a_{2l}$, $b_{2l}$) and odd ($a_{2l+1}$,
$b_{2l+1}$) indices. Therefore the determinant in the (\ref{gen_formula}) can be
written as a product of determinants corresponding to the even and odd amplitudes
\be
\Det\l[1-\l(\f{\hat{N}-1}{ \hat{N}+b}\,D\r)^2\r]=\Det\l[1-\l(\f{2 \, 
\hat{N}-1}{\hat{2 \, N}+b}\,U\r)^2\r]\,
\Det\l[1-\l( \f{2 \, \hat{N}}{2 \, \hat{N}+1+b}\,V\r)^2\r],
\label{uvd}
\ee
where the matrix elements of the matrices $U$ and $V$ are
\be
U_{pk}=  d_{2p\,2k} \,,~~~~~
V_{pk}=d_{2p+1,\,2k+1}~,
\ee
and the indices $p$ and $k$ take values $1,2,\ldots$\,.

For practical calculations it is also convenient to write the expressions for
the elements of the matrices entering in Eq.(\ref{uvd}) in terms of their
indices:
\bea
\l(\f{2 \,  \hat{N}-1}{\hat{2 \, N}+b}\,U\r)_{pk} &=& 2 \, {2 p - 1 \over 2 p
+b} \, {(2p+2 k -1)! \over (2p)! \, (2 k-1)!} \, \zeta(2p+2k) \, \left ( {R
\over \beta} \right )^{2p+2k}~, \nonumber \\
-\l( \f{2 \, \hat{N}}{2 \, \hat{N}+1+b}\,V\r)_{pk} &=& {4 p \over 2p+1 +b} \,
{(2p+2k+1)! \over (2p+1)! \, (2 k)!} \, \zeta(2p+2k+2) \, \left ( {R \over
\beta} \right )^{2p+2k+2}~,
\label{uvme}
\eea
with $p,k=1,2,\ldots$.

In order to find the first thermal correction at low temperature one can  expand
the matrices in power series using well known formula for the determinant
\be
\mathrm{Det}(1-M)=\exp\l[\mathrm{Tr}\ln\l(1-M\r)\r]=\exp\l[-\mathrm{Tr}\sum_{l=1
}^\infty\f{M^l}{l}\r]=1
-\mathrm{Tr}M+O(M^2).
\ee
In our case
\be
\l(\f{2 \, \hat{N}-1}{2 \, \hat{N}+b}\,U\r)^2=\l(
\ba{ccc}
\dst
\f{36\,\zeta^2(4)}{(2+b)^2}\l(\f{R _ s}{\b}\r)^8+\f{600\,\zeta(6)^2}{(2+b)\,(4+b)}\,
\l(\f{R _ s}{\b}\r)^{12} & \dst
\f{120\,\zeta(4)\zeta(6)}{(2+b)^2}\,\l(\f{R _ s}{\b}\r)^{10} & \cdots \\ \dst
\f{180\,\zeta(4)\zeta(6)}{(2+b)\,(4+b)}\,\l(\f{R _ s}{\b}\r)^{10} &
\dst\f{600\,\zeta(6)^2}{(2+b)\,(4+b)}\,\l(\f{R _ s}{\b}\r)^{12} & \cdots \\
\vdots & \vdots & \ddots
\ea
\r)~,
\ee
\be
\l(\f{2 \, \hat{N}}{2\, \hat{N}+1+b}\,V\r)^2=\l(
\ba{cc}
\dst \dst\f{1600\,\zeta(6)^2}{(3+b)^2}\,\l(\f{R _ s}{\b}\r)^{12} & \cdots \\
\vdots & \ddots
\ea
\r)~.
\ee
Therefore the first correction to the zero temperature value of the rate is
proportional to $R _ s^8/\b^8$ and is given by Eq.(\ref{kf}).

\begin{figure}[ht]
  \begin{center}
    \leavevmode
    \epsfxsize=7cm
    \epsfbox{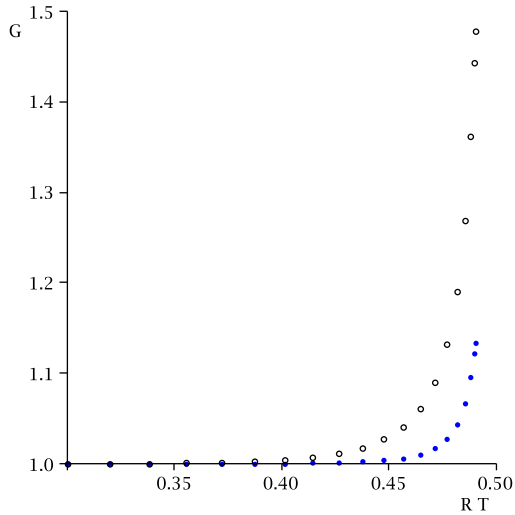}
    \caption{The thermal catalysis factor per each transverse dimension vs $RT$
for $b=1$ (open circles) and for $b=10$ (solid circles)}
  \end{center}
\label{corrections}
\end{figure}
The series for the function $G(T)$ diverges when $\b=2\,R _ s$. The corrections for
the temperature close to $(2R _ s)^{-1}$ can be found numerically. The proximity to
this point defines the number of terms which should be taken into account in the
series for $G(T)$. The plot for the function $G(T)$ vs the parameter $R _ s/\b=R _ s\,T$
calculated numerically with 50 first rows and columns retained in each of the matrices
$U$ and $V$ is shown in Fig.~8.

%\newpage
\subsection{Decay of a string induced by two-particle collisions}
As we have seen in the previous sections, the only difference between the calculation of the decay at finite temperature
$T$ and at $T=0$ arises at the level of calculating the pre-exponential factor
due to the functional determinant of the quadratic part of the action and
amounts to the standard treatment of the boundary conditions in (Euclidean) time
for the fluctuations: zero boundary conditions at large time for considering
zero temperature and periodic boundary conditions with period $\beta=1/T$ at
finite temperature.

The formula (\ref{rz}) is in  fact the unitarity relation\,\cite{Stone}
between the decay rate and the imaginary part of the amplitude of the transition
amplitude from the false vacuum to the false vacuum $\langle ({\rm
vacuum~1})_{\rm out}\,| ({\rm vacuum~1})_{\rm in} \rangle$. Similarly, one can
treat the probability of the string breakup by an excited state in terms of the
bounce contribution to the imaginary part of its forward scattering amplitude.
Proceeding to discussion of the string decay induced by the Goldstone bosons, we
readily notice that a state with just one Goldstone boson cannot induce the
destruction of the string. Indeed the total probability of such induced process
is  Lorentz invariant and thus can depend only on the (Lorentz) square of the
particle momentum $k^2$. The Goldstone bosons are massless, so that for them
$k^2=0$ and is fixed. If a single massless boson produced an effect on the
decay, this effect would thus be not depend on the energy $\omega$ of the boson.
In the limit $\omega \to 0$ the Goldstone boson is indistinguishable from the
vacuum. (In other words the limit $\omega \to 0$ corresponds to an overall shift
of the string in transverse direction.) Thus the decay rate of a single - boson
state is the same as that of the vacuum, and the presence of a single boson
with any energy produces no effect.

The simplest excited state, contributing to the string destruction, is that with
two particles. The probability $W_2$ of creation of the critical
gap in a  collision of two particles with
the two-momenta $k_1$ and $k_2$ can depend only on the Lorentz invariant
$s=(k_1+k_2)^2$. Obviously, for two particles colliding on a string $s=4
\omega_1 \omega_2$ (and $s=0$ for two particles moving in the same direction,
i.e. non-colliding). Using the unitarity relation, this probability can in principle be found
in terms of the imaginary part of the forward scattering amplitude
$A(k_1,k_2;k_1,k_2)$:
\be
W_2 = C { {\rm Im} A(k_1,k_2;k_1,k_2) \over \omega_1 \omega_2}~
\label{w2u}
\ee
where the factor $\omega_1 \omega_2$ is the usual flux factor, and the constant
$C$ does not depend on either of the energies, and is determined by specific
convention on the normalization of the amplitude.

The dynamics of the Goldstone bosons on the string, including their scattering,
can be considered in terms of the transverse shift $z_i(x)$ treated as a
two-dimensional field  described by the Nambu-Goto action (\ref{a0}) as
\be
S=\e_1 \, \int_{A_1} \, \sqrt{1+ (\partial_\a z_i)^2} \, d^2x +  \e_2 \,
\int_{A_2} \, \sqrt{1+ (\partial_\a z_i)^2} \, d^2 x + \m \int_P \sqrt{1+
(\partial z_i/\partial l)^2} \, dl~,
\label{ngg}
\ee
where $dl$ is the element of the length of the interface $P$ between the phases.

Clearly, at low energy of the Goldstone bosons one can make use of the expansion
in Eq.(\ref{ngg}) in powers of $(\partial z)$ which generates the expansion of
the scattering amplitudes in the momenta of the particles with each one entering
the amplitude with (at least) one power of its energy $\omega$, as is mandatory
for Goldstone bosons. For the scattering in the metastable state this generates
expansion in powers of $\omega / \sqrt{\e_1}$, so that in the lowest order in
this ratio, that we are discussing here, it is sufficient to retain only the
quadratic in $(\partial z)$ terms in the action (\ref{ngg}). It should be noted
that in spite of retaining only the quadratic terms, the multi-boson scattering
amplitudes do not vanish, since the necessary non-linearity arises from the
bounce configuration. In other words, the bosons scatter `through the bounce'.
The energy expansion for these amplitudes is determined by the bounce scale
$\ell_c$, so that the parameter of such expansion is $(\omega  \ell_c)$ which we
do not assume to be small. Clearly,  the condition for applicability of the
present approach where the terms of order $\omega / \sqrt{\e_1}$ are dropped,
while those with the parameter $\omega \ell_c$ are retained is that
\be
{\m^2 \over \e_1-\e_2} \gg 1-{\e_2 \over \e_1}~,
\label{condr}
\ee
which is always true if the semiclassical tunneling can be applied at all to the
string decay.

We shall now show that in the on-shell scattering through the bounce each
external leg enters with at least two powers of its energy. Let us start, for
the simplicity of illustration, with the binary scattering. The general two
$\to$ two scattering amplitude $ A(k_1,k_2;k_3,k_4)$ can be related in the
standard application of the reduction formula to the connected 4-point Green's
function $\langle {\rm vacuum~1}| T\{z(x_1) z(x_2) z(x_3) z(x_4) \}| {\rm
vacuum~1} \rangle$, which in turn is an analytical continuation of the Euclidean
connected correlator
\be
\langle {\rm vacuum~1}| z(x_1) z(x_2) z(x_3) z(x_4) | {\rm vacuum~1} \rangle =
{\cal Z}_0^{-1} \, {\delta^4 \, {\cal Z}_b[j] \over \delta j(x_1)\, \delta
j(x_2) \, \delta j(x_3) \, \delta j(x_4) } \left . \right |_{j=0}~.
\label{zj}
\ee
The latter expression for the correlator assumes the conventional
procedure of introducing in the action the source term $\int j(x) \, z(x) \,
d^2x$ for the Goldstone variable $z(x)$ and ${\cal Z}_b[j]$ is
the path integral around the bounce in the presence of the source.

\begin{figure}[ht]
  \begin{center}
    \leavevmode
    \epsfxsize=12cm
    \epsfbox{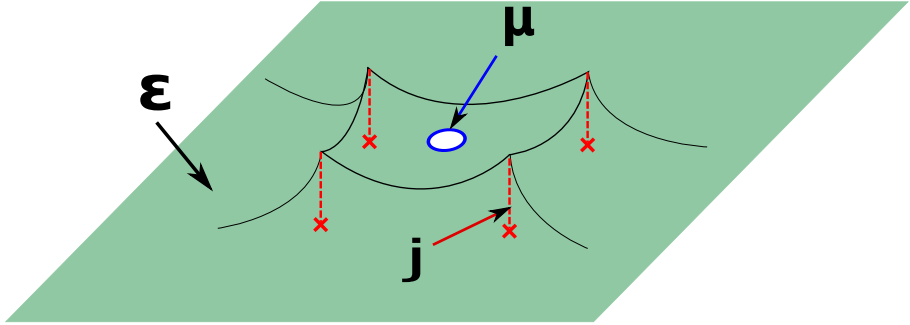}
    \caption{Bounce configuration distorted by the sources}
  \end{center}
\label{4_source}
\end{figure}

The low-energy limit of the on-shell scattering amplitude is determined by the
correlator at widely separated points $x_1,\ldots,x_4$. Weak $\delta$-function
sources `prop' the string in the transverse direction at those points as shown
in Fig.~9, and generally distort the bounce located between those points.   The
correlator (\ref{zj}) is determined by the distortion of the bounce by all four
sources, so that at large separation between the points the bounce is located
far (in the scale of its size $\ell_c$) from the sources, where the overall
distortion of the world sheet for the string is small and is slowly varying on
the scale $\ell_c$. One can therefore expand the background field $z_s$
generated by the sources at the bounce location in the Taylor series around an
arbitrarily chosen inside the bounce point $x=0$. Clearly the first term of this
expansion $z_s(0)$ corresponds to an overall transverse shift of the string and
does not alter the bounce shape and the action. Moreover the linear term in this
expansion, proportional to the gradient $\partial z_s(0)$ does not change the
bounce action either. Indeed this term corresponds to a linear incline of the
string in the transverse coordinates, and can be eliminated by an overall
rotation of the string in the $d$ dimensional space. In other words the absence
of the linear in the gradient of $z_s$ term in the action is a direct
consequence of the $d$-dimensional Lorentz invariance of the string. We thus
arrive at the conclusion that the expansion for the distortion of the bounce
starts from the second order in the derivatives of $z_s$ (the curvature of the
background world sheet), which for a connected correlator implies that in the
expansion in the energy each external source enters with at least two powers of
the energy. This conclusion clearly applies to the on-shell amplitudes with
arbitrary number of external legs, since the generalization to multiparticle
scattering is straightforward.

In particular, the binary scattering amplitude $ A(k_1,k_2;k_3,k_4)$ at low
energy scales as the eighth power of the energy, and the related to it (by the
Eq.(\ref{w2u})) probability of the string destruction by collision of two
particles is proportional to the sixth power of the energy scale, or
equivalently, to $s^3$ at small $s$. Applying in the same manner the unitarity
condition to the forward scattering amplitude of a general $n$-particle state,
one readily concludes that the corresponding probability $W_n$ of the induced
string breakup scales with the overall energy scale $\omega$ as  $W_n \propto
\omega^{3n}$.

\subsubsection{Thermal decay and the string destruction by particle collisions}

The described procedure for calculating the bounce-induced scattering amplitudes
in terms of the Euclidean correlators runs into the technical difficulty of
calculating the bounce distortion by the background created by the sources.
Furthermore, this procedure obviously involves a great redundancy, if the final
purpose is to calculate the probabilities $W_n$, i.e. only the absorptive parts
of the forward scattering on-shell amplitudes are the quantities of interest. We
find that in fact one can directly calculate the probabilities $W_n$ by an
appropriate interpretation of the more readily calculable thermal decay rate in
terms of the collision-induced probabilities. Such an approach is technically
more tractable due to the fact that in the thermal calculation the bounce is not
distorted, i.e. it is still a flat disk, and only the boundary conditions for
the modes of fluctuations are modified. We first illustrate this approach by
using the first nontrivial term of the low temperature expansion in
Eq.(\ref{klow}) for calculation of the low energy limit of the binary
probability $W_2$.

Indeed, the temperature dependent factors in the catalysis factor ${\cal K}_s$ arise from
the contribution to the critical gap nucleation of the boson collision, weighed
with the thermal number density distribution for the massless Goldstone bosons
\be
d\,n(k) = {1 \over e^{\omega/T} - 1} \, {d \, k \over 2 \pi}~,
\label{bose}
\ee
where $k$ is the spatial momentum, $|k|=\omega$, running from $-\infty$ to
$+\infty$. Given the low energy behavior $W_n \propto \omega^{3n}$ found in the
previous section, one readily concludes that the contribution of $n$-particle
string destruction starts with the term $T^{4n}$ in the low $T$ expansion for
$K$. Thus the first term written in Eq.(\ref{klow}) can arise only from $n=2$
i.e. from the binary production of the critical gap. Writing the expansion in
$s$ of the probability $W_2$ as $W_2=c_3 \, s^3  + \ldots$, one determines the
coefficient $c_3$ of the $s^3$ term  by comparing the result in Eq.(\ref{klow})
with the one calculated in terms of the two-particle collision rate using the
number density distribution (\ref{bose}):
\bea
&&(d-2) \, \gamma _ s\, {\pi^8 \over 450}\, \left ( {\e_1-\e_2 \over 3 \e_1-\e_2}
\right
)^2 \, \l(\f{\ell_c T}{2}\r)^8 = \nonumber \\
&&(d-2) \, c_3 \, \int_0^\infty {d\omega_1 \over  e^{\omega_1/T} - 1} \,
\int_0^\infty {d \omega_2 \over  e^{\omega_2/T} - 1}\, {s^3 \over 4 \pi^2}
 = (d-2) \, c_3 \, {16 \, \pi^6 \over 225} \,  T^8~,
\label{w2low}
\eea
where the relation $s=4 \omega_1 \omega_2$ is used and the integration is done
only for the bosons with opposite signs of their momenta, and avoiding the
double-counting for the identical particles. The factor $(d-2)$ counting the
number of the transverse dimension, corresponds to the summation over the
polarizations of the Goldstone bosons. The expression for the coefficient $c_3$
following from Eq.(\ref{w2low}) thus determines the first term in the expansion
for $W_2$
\be
W_2=\gamma_s \, R _ s ^ 2 \, \left [ { \pi^2 \over 32}  \, \left ( {\e_1-\e_2 \over 3
\e_1-\e_2} \right
)^2 \, R_s^6 \, s^3 + \ldots \right ]~.
\label{w2l1}
\ee 
(One can notice that at $\e_2=0$ the $s^3$ term coincides with the first term of
expansion of the expression in Eq.(\ref{w20}).)

\subsubsection{Thermal bath with a chemical potential \label{sec_therm_bath}}

The discussed procedure for extracting the coefficients of the energy expansion
for the probability of collision-induced string decay is obviously limited to
only the first term in $W_2$. In the higher terms in the temperature expansion
for ${\cal K}_s$ the contribution of the energy expansion for $W_n$ with different $n$
generally gets entangled. This happens because the temperature is the only
parameter and terms originating from different $n$ can contribute in the same
power in $T$. In order to disentangle the terms of higher order in energy in
$W_n$ with low $n$ from similar terms originating from higher $n$ we introduce a
{\em negative} chemical potential $\nu$ for the Goldstone bosons. Generally such
procedure would be impossible, since the number of these bosons is not
conserved. However in our application this procedure is fully legitimate. Indeed
the thermal state of the string that we study here is that of {\em
collisionless} bosons, in which their number {\em is} conserved. The string
decay, resulting in a change in this number, is a very weak process that we
consider only in the first order, which justifies averaging the rate of this
process over the unperturbed state with conserved number of particles. At
negative $\nu$ the number density distribution of the bosons (\ref{bose}) is
replaced by
\be
d \, n (k) = {1 \over e^{\omega + |\nu| \over T} - 1} \, {d \, k \over 2 \pi}~,
\label{bosem}
\ee
and by tuning the parameter $|\nu|/T$ one can readily resolve between the
contribution of $n$-particle processes with different $n$.

The introduction of the chemical potential requires us to modify our previous
thermal calculation (see Sec. \ref{sect2}). The modification of this calculation for a thermal state with a negative
chemical potential, where the number density of the bosons be given by
Eq.(\ref{bosem}), is achieved by introducing a `damping factor' in the periodic
sums for the outer functions in Eq.(\ref{syst}):
\be
g_k(w) \to g^{(\nu)}_k(w)
=\f{R _ s^k}{w^k}+\sum_{n=1}^\infty\l[\f{R _ s^k}{(w-n\b)^k}+\f{R _ s^k}{(w+n\b)^k}\r] \,
\exp \left( - n \, |\nu| \, \beta \right ) ~.
\label{gkm}
\ee
One can readily see that the only net result of the $\nu$ dependent factor in
the calculation is a modification of the matrix coefficients
$d_{pk}$ amounting to the replacement of the factors $\zeta(q)$ by the standard
polylogarithm function,
$$\Li_q(x)=\sum_{n=1}^\infty { x^n \over n^q}~,$$
$\zeta(q) \to \Li_q(e^{-|\nu|/T})$. In other words, the catalysis factor for a
thermal state with a negative chemical potential is given by the expression
\be
\mathcal { K }  _ s (\nu,T) = \Det\l[1-{\cal D}^2(\nu, T)\r]^{-(d-2)/2}~,
\label{kmt}
\ee
with the elements of the matrix ${\cal D}(\nu, T)$ having the form
\be
\left [ {\cal D}(\nu,T) \right ]_{pk}= -\l[(-1)^k+(-1)^p\r] \l ( R _ s T
\r)^{k+p+2}\,{p \over p+b} \, {(p+k+1)! \over (p+1)! \, k! } \, \Li_{p+k+2}
\left ( e^{-|\nu|/T} \right )~.
\label{cdmt}
\ee

The dependence on two `tunable' parameters $\nu$ and $T$ in Eq.(\ref{kmt}) makes
it possible to disentangle the contribution of processes with different number
of particles from the energy behavior in each of these processes. Such a
separation becomes straightforward if one notices that each factor with the
polylogarithm $\Li$ arises from the integration over the distribution
(\ref{bosem}):
\be
\int_0^\infty {\omega^q \over e^{\omega + |\nu| \over T} - 1} \, d \omega = q!
\, T^{q+1} \, \Li_{q+1} \left ( e^{-|\nu|/T} \right )~.
\label{bosint}
\ee
One therefore concludes that the number of the `$Li$ factors' in each term of
the expansion of the catalysis factor in Eq.(\ref{kmt}) directly gives the
number of particles in the process, while the indices of these polylogarithmic
factors give the power of the energy for each particle. Given that the matrix
${\cal D}(\nu,T)$ is linear in the `$Li$ factors', one can count the number of
particles contributing to each term of the expansion for $\mathcal{ K(\nu,T) }_ s$ by counting
instead the power of  ${\cal D}(\nu,T)$. The latter counting is simplified if
one rewrites the equation (\ref{kmt}) in the equivalent form, suitable for the
expansion in powers of ${\cal D}(\nu,T)$:
\bea
\label{dexp}
\mathcal { K }  _ s (\nu,T) &=& \exp \left \{ - {d-2 \over 2} \, \Tr  \ln \left [1- {\cal
D}(\nu,T)^2 \right ] \right \} = \\  \nonumber
1+{d-2 \over 2} \, \Tr \left [ {\cal D}(\nu,T)^2 \right ] & + & {d-2 \over 4} \,
\Tr \left [ {\cal D}(\nu,T)^4 \right ] + {(d-2)^2 \over 8} \, \left \{ \Tr \left
[ {\cal D}(\nu,T)^2 \right ] \right \}^2 + O \left ( {\cal D}^6 \right ).
\eea

The latter expression merits some observations. The first is that all the terms
in the expansion in powers of ${\cal D}(\nu,T)$ are positive, which is certainly
the necessary condition for the consistency of our interpretation of these terms
as corresponding to the probability of the destruction of the string by
$n$-particle collisions. The second is that the string is destroyed only in
collisions of {\em even} number of particles, since the expansion goes in the
even powers of ${\cal D}(\nu,T)$. Finally, the third observation is related to
the dependence in Eq.(\ref{dexp}) on the number of the transverse dimensions
$(d-2)$. Namely, the quadratic in ${\cal D}(\nu,T)$ term is proportional to
$(d-2)$. This corresponds to that in two-particle collisions only the Goldstone
bosons with the same transverse polarization do destroy the string. On the
contrary, the quartic in ${\cal D}(\nu,T)$ term has one contribution
proportional to $(d-2)$ and one proportional to $(d-2)^2$. The linear in $(d-2)$
part  corresponds to  all the bosons in the collision having the same
polarization, while the quadratic in $(d-2)$ part is necessarily contributed by
the collisions, where the colliding bosons have different polarization.

\subsubsection{Destruction of the string in two-particle collisions at arbitrary $
R_s \, \sqrt{s}$}

The expression (\ref{dexp}) for the catalysis factor $\mathcal { K }  _ s (\nu,T)$ together with
the formulas (\ref{cdmt}) and (\ref{bosint}) reduce the calculation of the
probability of the string breakup by a collision of an arbitrary (even) number
$n$ of particles to straightforward, although not necessarily short, algebraic
manipulations. In this section we consider in full the most physically
transparent case of two-particle collisions. The probability in this case is
found from the term in Eq.(\ref{dexp}) with the trace $\Tr  [ {\cal D}(\nu,T)^2 
]$. Using Eq.(\ref{cdmt}), this trace can be written as a double sum:
\bea
\label{trd2}
&&\Tr \left [ {\cal D}(\nu,T)^2 \right ] = \\
\nonumber
&&4 \, \sum_{p=1}^\infty \sum_{k=1}^\infty \, { 1 \over (p+b+1) \, (k+b+1) } \,
{\left[ (p+k+1)! \,(R _ s T)^{p+k+2} \, \Li_{p+k+2} \left ( e^{-|\nu|/T} \right )
\right ]^2  \over (p-1)! \, (p+1)! \, (k-1)! \, (k+1)! }
\eea
One can readily recognize the expression in the straight braces here as the
integral (\ref{bosint}) with the power of the energy $q$ given by $(p+k+1)$, and
thus identify the coefficient of the same power of $s=4 \omega_1 \omega_2$ in
the expansion of the probability $W_2(s)$ in $s$. In this way we find the
following formula for $W_2(s)$ in terms of this expansion,
\bea
W_2(s) & = & 8 \, \pi^2 \, \gamma _ s \, R _ s ^2 \,  \sum_{p=1}^\infty
\sum_{k=1}^\infty \, { 1 \over (p+b+1) \, (k+b+1) } \, {(s R _ s ^ 2 / 4)^{p+k+1}  \over
(p-1)! \, (p+1)! \, (k-1)! \, (k+1)! } \nonumber \\
&=& 8 \, \pi^2 \, \gamma _ s \, R _ s ^2 \,  \left [ \Phi_b(\sqrt{s} \, R _ s ) \right ]^2~,
\label{w2fin}
\eea
where the function $\Phi_b(x)$ expands in a single series as
\be
\Phi_b (x) = { x \over 2} \sum_{p=1}^\infty {1 \over p+b+1} \, {x^{2p} \over
(p-1)! \, (p+1)! }~.
\label{phib}
\ee
It can be noted that the two-particle probability depends only on the odd powers
of $s$. This in fact is consequence of the binary forward scattering amplitude
being even in $s$, as required by the Bose symmetry.

For integer values of the parameter $b$, the function $\Phi_b(x)$ has a simple
expression in terms of the modified Bessel functions $I_q(x)$ of the order $q$
up to $q=b+2$. This expression is especially simple for $b=1$, i.e. for the case
of the string decay into `nothing': $\Phi_1(x)=I_3(x)$, so that one arrives at
the formula (\ref{w20}). We also write here, for an illustration, the
corresponding expressions for the next two integer values of $b$:
\be
\Phi_2(x)={1 \over x} \, \left [ I_5(x) + 6 I_4(x) \right]\,;~~~~~~~
\Phi_3(x)={1 \over x^2} \, \left [ (48+ x^2) \, I_5(x) + x \, I_6(x) \right]\,~.
\label{phibex}
\ee

%\newpage
\section{Domain wall}
\subsection{Spontaneous decay of a metastable domain wall}
In the following sections we  generalize the previously obtained results for the decay of a metastable domain wall. We denote the tension of a domain wall by $\ep$ and the tension of a string associated with the edge of the wall by $\s$, so that there should not be any confusion with previously used notations for a string. For a complete decay of a domain wall the low energy effective action is similar to (\ref{a0}) and is given by Nambu-Goto
action for two and tree dimensional objects
\be
S = \s \, \mathcal {A} + \ep \, \mathcal { V },
\label{NG}
\ee
with $\mathcal { V }$ being the world volume of the wall, while $\mathcal {A}$ is the world
area of interface. As before this action does not take into account the inner structure of the wall or the interface. A nontrivial classical solution (bounce) in this case is empty sphere with the radius
\be
R _ w = \f {2 \s} {\ep},
\ee
surrounded by the metastable phase (Fig.~10).
\begin{figure}[ht]
  \begin{center}
    \leavevmode
    \epsfxsize=7cm
    \epsfbox{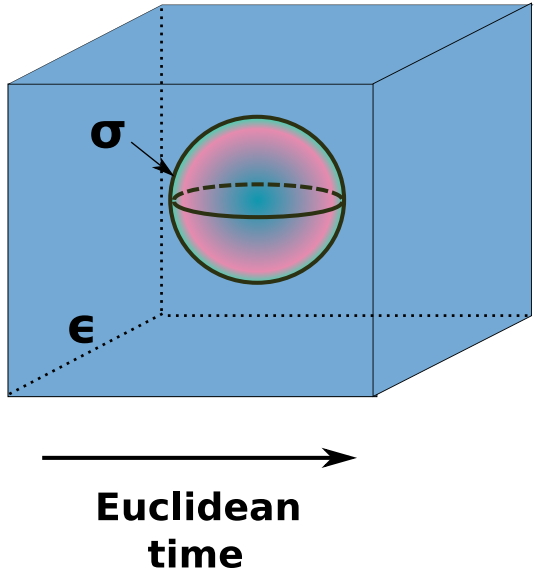}
    \caption{Bounce configuration}
  \end{center}
\label{wall_classical}
\end{figure}
The appropriate quantity to be found is $\gamma _ w $: the nucleation rate of critical holes per unit area (not length as it was for a string transition).

Following the same steps as in calculating the decay rate of a string we find\footnote{The factor ${\mathcal{C} \over \ep ^ {7/3} } \, \exp \l( - {16 \, \pi \,
\mu^3 \over 3 \, \eps^2} \r) $ corresponds to the decay rate of a metastable vacuum in $2+1$ dimensions \cite{mv2004}.}
\be
\gamma _w = {\mathcal{C} \over \ep ^ {7/3} } \, \exp \l( - {16 \, \pi \,
\s^3 \over 3 \, \ep^2} \r) \,
{{{\cal Z}}_{12} \over {\cal
Z}_1}~,
\ee
with boundary partition functions given by
\be
\mathcal{Z}_ {12 (\rm boundary)} = \int {\cal D} \x \, \exp \l[ - {\s \over 2}
\, \int d \Omega \, \x \, \Delta ^ { (2) } \x + {\ep R _ w ^ 2 \over 2} \, \int d
\Omega \, \x \, \p_r z_c \Big | _ { r = R _ w } \r],
\label{z12b}
\ee
and
\be
\mathcal{Z}_ {1 (\rm boundary)} = \int {\cal D} \x \, \exp \l[{\ep R _ w ^ 2 \over 2}
\, \int d \Omega \, \x \, \p_r z_c \Big | _ { r = R } \r],
\label{z1b}
\ee
with the function $z_c(r,\t,\vp)$ satisfying the Laplace equation $\Delta \,
z_c=0$ with the boundary conditions
\be
z _c (R _ w, \t, \vp) = \x (\t, \vp), ~~~~ z _c (r = L)=0.
\label{zcbc_w}
\ee
The operator $\Delta ^ { (2) }$ is the angular part of the Laplace operator in
$3$d (the Laplace operator on a sphere).
Expanding the boundary
function $\x (\t, \vp)$ in the series in spherical harmonics
\be
\x(\t, \vp) = \sum _ {l , m} A _ {l m} Y _ {l m} (\t, \vp) ~ ,
\label{xiw}
\ee
we find a complete set of functions
\be
z_{lm}(r, \t, \vp) = A _ {l m} \f {R _ w ^ {l+1}} {r ^ {l+1}} \, Y _ {l m} (\t, \vp) ~
.
\label{zlm0}
\ee
Substituting these functions to the equations (\ref{z12b}) and (\ref{z1b}) and
performing integration over the amplitudes
$A _ {l m}$ yields
\be
\mathcal{Z}_ {12 (\rm boundary)} = {\cal N} \prod _ { l=0 } ^ { \infty }
\l( \f {1} {\s \, l\, (l+1) + \ep \, R_w\, (l+1)} \r) ^ { (2 l + 1) /2 }
\ee
and
\be
\mathcal{Z}_ {1 (\rm boundary)} = {\cal N} \prod _ { l=0 } ^ { \infty }
\l( \f {1} {\ep \, R_w\, (l+1)} \r) ^ { (2 l + 1) /2 }.
\ee

\subsubsection{Regularization}
Each of the expressions (\ref{z12b}) and (\ref{z1b})
contains a divergent product (compare to Eqs.(\ref{z12by1}) and (\ref{z1by1})). As previously,  introducing Pauli-Villars regulators with masses $M_\a$, we find the ratio of the boundary partition functions to be equal to
\bea
\mathcal {R} _ w& = & \prod_{l=0}^\infty \l[ \f { l\, (l+1) + 2 \sqrt{M^2_\a \,R _ w ^ 2
+ \l(l+\f {1} {2} \r)^2} +
1} { (l+1) \, (l+2)} \r] ^ { {(2l+1) \, C_\a} / {2} } \times \nonumber \\
&& \prod_{l=0}^\infty \l[ \f {l + 1} { \sqrt{M^2_\a \,R _ w ^2 + \l(l+\f {1} {2}
\r)^2} +
\f {1} {2} } \r] ^ { {(2l+1) \, C_\a} / {2} } \times
\nonumber \\
&& \prod_{l=0}^\infty \l[ 1 + \f {M_ \a ^2 R _ w ^2} { \l( M^2_\a \,R _ w ^2 + \l(l+\f {1}
{2} \r)^2 \r) \,
\l( l\, (l+1) + 2 \sqrt{M^2_\a \,R _ w ^2 + \l(l+\f {1} {2} \r)^2} + 1 \r)} \r] ^ {
{(2l+1) \, C_\a} / {2} } ~,
\label{crprod_w}
\eea
where we took into account that $R  _ w = 2 \, \s / \ep$.

Now, each of the products in Eq.(\ref{crprod}) is finite at a finite $M_\a$ and can be
calculated separately. Instead of calculating the product directly, we can use
the relation
\be
\prod_l F_l = \exp \l( \sum_l \ln F_l \r),
\ee
and calculate the sum. We start from the third product. The expression under the
sign of product is of the form
$1+g(l)$, with $g(l)$ close to $0$ for any $l$, since it behaves as $M^{-1}_ {
\a }$. Hence we can leave only the first term in the expansion of the logarithm
$\ln ( 1+x ) = x + O (x^2)$. Thus, we need to find the sum
\be
S_3 = \f {1} {2} \sum_ {l = 0} ^ \infty \sum_ {\a} C_ \a \, (2l+1) \, {M_ \a ^2
R _ w ^2} { \l[ M^2_\a \,R _ w ^2 + \l(l+\f {1} {2} \r)^2 \r] ^ {-1} \,
\l[ l\, (l+1) + 2 \sqrt{M^2_\a \,R _ w ^2 + \l(l+\f {1} {2} \r)^2} + 1 \r] ^ {-1}}~.
\ee
The sums associated with the three products are readily calculable with the help
of the Euler-Maclaurin summation formula and  the result for the regularized
ratio has the form
\be
\mathcal{R} _ w = \exp \l[ \f {1} {2} M^2 R _ w ^2 \ln M R _ w + M R _ w \ln M R _ w + \ln M R _ w \r]
\label{reg_r_M}
\ee
where $M^n \ln M = \sum _\a C _\a M ^n _\a \ln M _\a$, for any $n$. The
expression for $\mathcal{R}_w$ contains an essential dependence on the regulator
mass parameter $M$. However, similarly to the previously considered case all such dependence in the phase
transition rate can be absorbed in renormalization of the parameter $\s$ in the
leading semiclassical term. 

Using the same technique as employed in Section \ref{sect1} one can find the renormalized parameter $\s$ to be given by
\be
\s_R = \s + \delta \s
\ee
with
\bea
\delta \s & = & \f {1} {2} \int \f {d^2\, k} {(2 \pi)^2} \l[ \ln \l( k^2 + \f
{\e} {\m} \sqrt {M^2 + k^2} \r) -
\ln \l( k^2 + \f {\ep} {\s} \sqrt {k^2} \r) \r] - \nonumber \\
&& \f {1} {4} \int \f {d^2\, k} {(2 \pi)^2} \l[ \ln \l( M^2 + k^2 \r) -
\ln k^2 \r] = \nonumber \\
&& \f {M^2} {8 \pi} \ln M R _ w + \f {M \e} {8 \pi \m} \ln M R _ w + \f {\e^2} {16
\pi\m^2} \ln M R _ w.
\label{reg_m}
\eea

\subsubsection{Results}
Collecting all terms together we find the rate of the process. It is clear that
the result for each $d-3$ transverse dimensions factorizes, thus we have the
expression for the rate in the form
\be
\gamma _ w = \f {\mathcal {C}} {\ep ^ {7/3}} \, \mathcal{R} _ w ^ {d-3} \,
\exp \l( - {16 \, \pi \, \s ^ 3 \over 3 \, \ep ^ 2 } \r) ,
\ee
where $\s$ is the bare (non-renormalized) tension, and the regularized ratio
$\mathcal{R}$ is given by (\ref{reg_r_M}). Taking into account that each of the
transverse dimensions contributes additively to $\delta \s$ and the interface
is a sphere with area
\be
\mathcal{A} = 4 \pi R _ w ^ 2,
\ee
and expressing the bare $\s$ through the renormalized one: $\s=\s _ R-\delta
\s$, one readily finds that the dependence on the regulator mass M cancels in
the transition rate, and one arrives at the formula given by Eq.(\ref{rate}).

Thus we obtained the result similar to the decay of a string, where the effect of
all extra transverse dimensions results only in the renormalization of parameter
$\s$ associated with the interface. It should be mentioned, however, that the
result for a string was, actually, obtained for a transition between two states
of a string with different tensions. Here we considered decay of a wall
(transition into nothing). For the calculation used it is impossible to preserve
finite terms, but only proportional to some power of the regulator mass
parameter $M$.

Having calculated the probability rate for a decay of one- and two- dimensional
objects, e.g. string and domain wall, it is tempting to assume that the
situation is somewhat similar for the decay of an object of arbitrary
dimensionality. But it is not true already for the decay of tree- and four-
dimensional objects, where the dependence of the result on regulator mass is
substantial even after renormalization of a parameter associated with an
interface. That dependence demands introduction of new terms into initial
action, which corresponds to nonrenormalizibility of the effective `low-energy'
theory.

\subsection{Thermally induced decay of a domain wall}

The thermal effects in decay of a metastable domain wall involve one important difference from those in the decay of a string. Namely, the previously considered expansion of the catalysis factor ${\cal K}_s$ is valid up to the temperature $T = 1/\ell_c=1/(2 R_s)$, at which point the whole calculation breaks down due to a change in the mechanism of the transition: the thermal fluctuations of the string start dominating over the quantum ones. Simultaneously, at this temperature the thermal factor in Eq.(\ref{gen_formula}) develops a singularity, and the low-temperature expansion diverges. In the case of a domain wall decay at low temperature the thermal effects in the quantum tunneling result in a catalysis factor ${\cal K}_w$ multiplying the semiclassical exponential factor in $\gamma_w$:
\be
\gamma_w(T)=\mathcal {K} _ w \gamma _w~.
\label{kwqu}
\ee
On the other hand, the static potential for a bubble, a round hole in the wall, depends on the radius $R$ of the bubble as
\be
V(R)=2 \pi \, \s \, R -  \pi \, \ep \, R^2~,
\label{wpotv}
\ee 
and this potential has a maximum at $R_m = \s/ \ep$ where its value is 
\be
V_m=V(R_m)={ \pi  \, \s^2 \over \ep}~.
\label{wpotvm}
\ee
The probability of classical thermal fluctuations over the barrier $V_m$ is proportional to the activation factor $\exp(-V_m/T)$, and this factor becomes larger than the exponential term in $\gamma_w$ (Eq.(\ref{rate})) starting from $T=3/(8 R_w)$, at which temperature the classical fluctuations over the barrier start to dominate over the considered here quantum tunneling. In terms of the Euclidean space formulation of the problem the periodic replication with the period $\beta$ of the spherical bounce (Fig.11a) gives a larger action per period\cite{Garriga94} than the cylindrical configuration in Fig.11b as soon as $T > 3/(8 R_w)$. For this reason our calculation of the thermal factor ${\cal K}_w$ makes physical sense only as long as the temperature is lower than this value. 
\begin{figure}[htb]
%    \leavevmode
 \begin{center}
  $\begin{array}{cc}
    {\epsfxsize=8cm \epsfbox{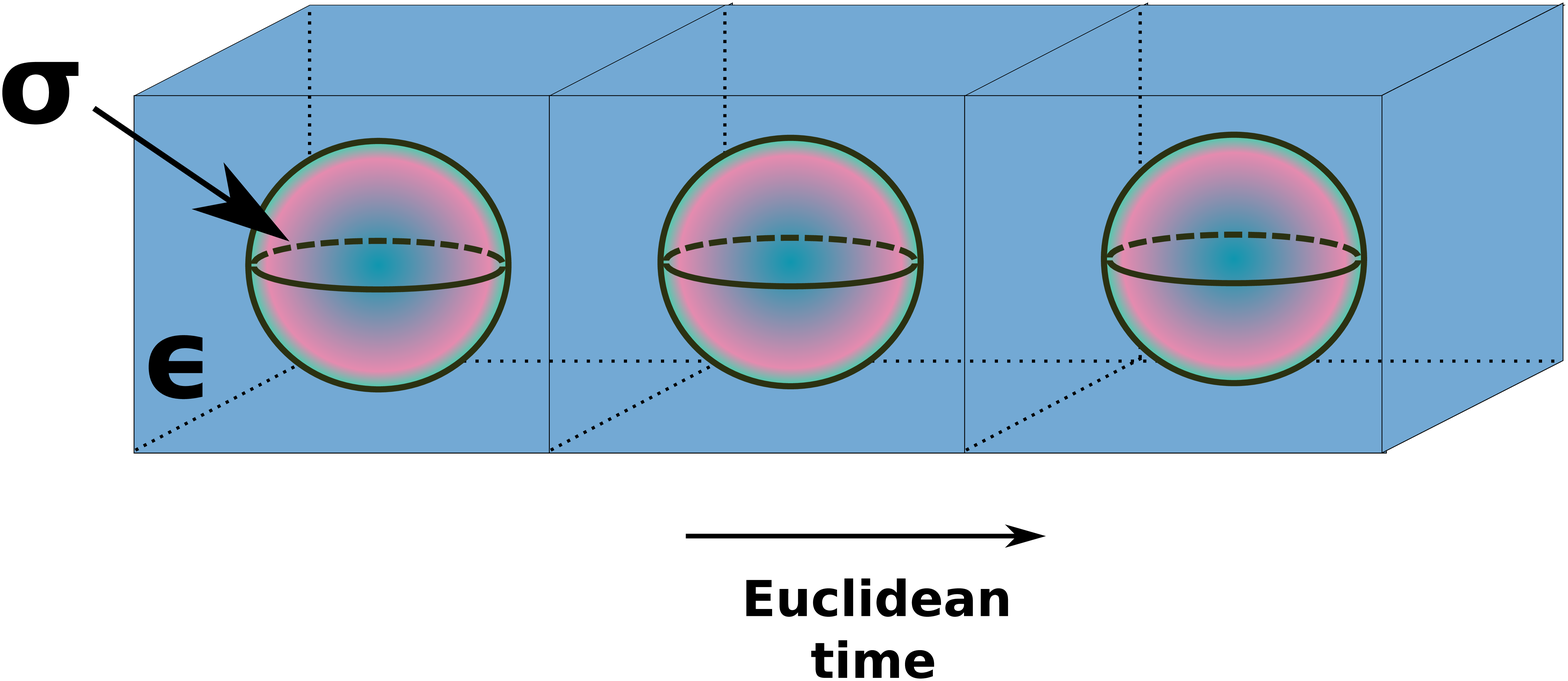}} & {\epsfxsize=8cm \epsfbox{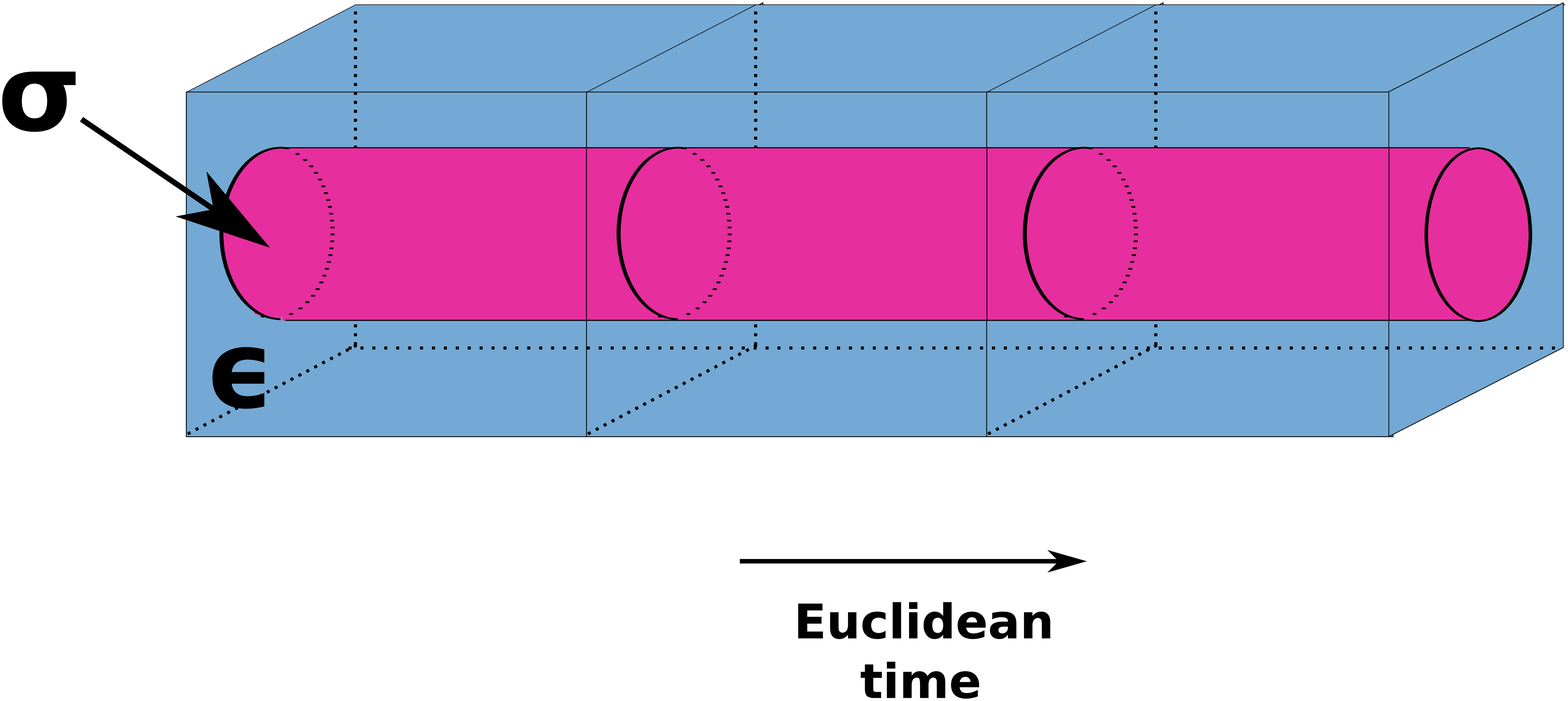}} \\
    (a) & (b)
   \end{array}$
   \caption{Bounce configuration for the quantum tunneling at low temperature (a) and for the thermally activated transition (b).}
\label{wall_classical_temperature}
\end{center}
\end{figure}

In calculating the preexponential factor arising in the periodically replicated bounce configuration,
we encounter the same problem of a mismatch between the symmetry of the boundary conditions and the symmetry of the bounce. However one can cope with this problem in a way similar to the previously discussed approach to calculating the decay rate of the string at finite temperature. Namely, one can consider the solution to the Laplace equation as one produced by the sources placed at each period (see Fig.~12). Thus the solution $z_{lm}$ (the periodic analog of the functions in Eq.(\ref{zlm0})) is given by
\be
z _ {l m} = \f {R _ w ^ { l + 1 } } {r  ^ { l + 1 }} Y _ { l m } (\t , \varphi) +
\sum _ { n \neq 0 } \f {R _ w ^ { l + 1 } } {r _ { n } ^ { l + 1 }} Y _ { l m } (\t _ { n } , \varphi),
\label{zlmT}
\ee
with $r _ n$ and $\t _ n$ reading as
\bea
r _ { n } & = &\sqrt { r ^ 2 + ( \b n ) ^ 2 - 2 \, \b n \, r  \cos \t }, \nonumber \\
\sin \t _ n & = & \f { r \sin \t } { r _ n }.
\label{thetasrs}
\eea
The solution (\ref{zlmT}) can be expanded in a power series
\bea
z _ {l m} &=& \f {R _ w ^ { l + 1 } } {r ^ { l + 1 }} Y _ { l m } (\t , \varphi) + \sum _ {l'} \sum _ { n > 0 } \f {R _ w ^ { l' } r ^ { l + 1 }}
{ ( n \b ) ^ { l + l' + 1 } } C _ {l l'} ^ { m } Y _ { l' m } (\t , \varphi) \nonumber \\
&=& \f {R _ w ^ { l + 1 } } {r ^ { l + 1 }} Y _ { l m } (\t , \varphi) + \sum _ {l'} \f {R _ w ^ { l' } r ^ { l + 1 }}
{ \b  ^ { l + l' + 1 } } \zeta (l + k + 1) C _ {l l'} ^ { m } Y _ { l' m } (\t , \varphi).
\label{zexpw}
\eea
\begin{figure}[ht]
  \begin{center}
    \leavevmode
    \epsfxsize=8cm
    \epsfbox{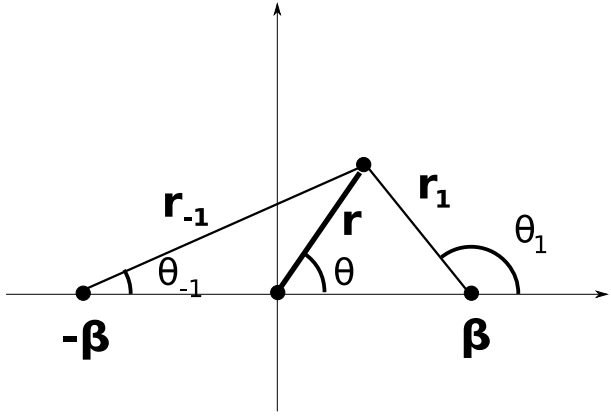}
    \caption{The illustration of the notation for the angles and distances used in Eq.(\ref{thetasrs}).}
  \end{center}
\label{periodic_wall}
\end{figure}

In order to find the constants $C _ {l l'} ^ { m }$ one can notice that the expression in Eq.(\ref{zlmT}) has the same structure for each $n$. Therefore, for the purposes at hand we can consider only two adjacent periods ($ \pm \b $). Then integrating the \textit{r.h.s.} of the (\ref{zlmT}) with $Y _ { l m } ^ {*} (\t , \varphi)$ we find the result to be
\be
C _ {l l'} ^ { m } = 2 (-1) ^ {l + l' + m} \sqrt { \f {2 l + 1} {2 l' + 1} } \f { ( l + l' ) !}
{ \sqrt { ( l + m )! \, ( l - m )! \, ( l' + m )! \, ( l' - m )! } }
\label{coeffc}
\ee 
It should be mentioned here that obviously
\be
\sum _ {m = - l} ^ {l} C _ {l l} ^ { m } =0.
\label{restr}
\ee
This relation can be understood even without actual calculation. Indeed, the expression
\be
\int \sum _ m Y _ { l m } ^ { * } (\t , \varphi) Y _ { l m } (\t _1, \varphi) \, \sin \t \, d \t \, d \varphi
\ee
is proportional to the electric potential at point $\b$ created by the set of $2 ^ l$-poles placed on a sphere with its center in the origin. Obviously, such a potential is equal to zero.

Once we have found the coefficients (\ref{coeffc}) of the expansion (\ref{zexpw}) we can follow the same routine as used in the section (\ref{sect2}). Namely, we can express any solution of the Laplace equation with periodic boundary conditions as a series 
\be
z _ {c} = \sum _ { l m } z _ {l m} B _ {l m}.
\label{zsol}
\ee
After that using the expansion of a boundary function (\ref{xiw}) and the relation (\ref{zcbc_w}), we express coefficients $B _ {l m}$ through $A _ {l m}$
\be
A _ {l m} = \l [\delta _ {l l'} + \l ( \f {R _ w} {\b} \r ) ^ {l + l' + 1} C _ { l l' } ^ { m } \, \zeta (l + l' + 1) \r ] \, B _ { l' m },
\ee
or
\be
A _ m = ( 1 + D _ m ) B _ m,
\label{ABrel}
\ee
where $D_m$ is a matrix with elements
\be
[D _ m] _ { l l'} = \l ( \f {R _ w} {\b} \r ) ^ {l + l' + 1} C _ { l l' } ^ { m } \, \zeta (l + l' + 1),
\label{dmatrw}
\ee
and $A _ m$ (similarly $B _ m$) is the row type object with elements
\be
[A _ m] _ { l } = A _ {l m}.
\ee
It should be mentioned that since the solution $z _ {lm}$ exists only for $l \geq m$, the row $A _ m$ has nonzero elements $A _ {l m}$ only for $l \geq m$.

Substituting the solution (\ref{zsol}) to the expressions for the partition functions (\ref{z12b}) and (\ref{z1b}) and integrating over the amplitudes $A _ {lm}$ (with a help of relation (\ref{ABrel})) we find the catalysis factor  in the following form
\be
\mathcal { K } _ w = \prod _ m \det \l (  1 + \f {N \, ( N - 1 )} { ( N + 1 ) \, ( N + 2 )} \, D _ m  \r ) ^ { - {{d-3}  \over  {2}} }, 
\ee
with matrix $N$ defined in (\ref{Nmatr}). Equivalently, one can rewrite the expression in the following form
\be
\mathcal { K } _ w = \det \l (  1 + \bar { \mathcal { D } } \r )^{ - {{d-3}  \over  {2}} },
\label{kwquf}
\ee
where $\mathcal { D }$ stands for the block diagonal matrix defined as
\be
\bar{\mathcal { D } }= \l (
\ba{llllll}
\mathcal { D } _ 0&&&&& \\
&\mathcal { D } _ 1&&&& \\
&&\ddots&&& \\
&&&\mathcal { D } _ m& \\
&&&&\ddots&
\ea
\r )
\ee
with elements
\be
\mathcal { D } _ m = \f {N \, ( N - 1 )} { ( N + 1 ) \, ( N + 2 )} \, D _ m.
\ee
The first nontrivial term of expansion of the catalysis factor $\mathcal { K } _ w$ at small temperatures is determined by the coefficients in Eq.~(\ref{dmatrw}) at $l=l'=2$ and the result for this term is given in  Eq.(\ref{kfwl}).

One can notice that the low-temperature expansion for the factor ${\cal K}_w$ generated by Eq.(\ref{kwquf}) is well behaved at $T R_w = 3/8$, beyond which temperature, as previously discussed, the regime of the transition changes from quantum tunneling to a classical thermal activation. Moreover, the calculated thermal effect at this point is numerically extremely small: ${\cal K}_w - 1 \sim 10^{-3}$. Thus the explicit form of higher terms in the temperature expansion is of a `practical' interest only inasmuch as the thermal calculation is used as a preliminary step for finding the generating function for the rate of the wall destruction in collisions of Goldstone bosons.

%\newpage
\subsection{Destruction of a metastable domain wall in binary collisions}
Using the arguments from the section \ref{sec_therm_bath} it is possible to relate the catalysis factor $\mathcal {K} _ w$ for a decay rate of the domain wall at finite temperature calculated in the previous section, to the effective length of a particle collision (the analog of the probability in $1+1$ dimensional case).

As before we consider the distribution function for the Goldstone bosons in the following form (with negative chemical potential $-|\nu|$)
\be
d n (\vec {k}) = \f {1} {e ^ { \f { \omega + |\nu| } {T} } - 1 } \f { d ^ 2 k} { ( 2 \pi ) ^ 2 }.
\ee
Such an introduction of the chemical potential modifies the result for the catalysis factor $\mathcal{K} _ w$ in the way that the solutions (\ref{zexpw}) are modified as
\bea
z _ {l m} &=& \f {R _ w ^ { l + 1 } } {r ^ { l + 1 }} Y _ { l m } (\t , \varphi) + \sum _ {l'} \sum _ { n > 0 } \f {R _ w ^ { l' } r ^ { l + 1 }}
{ ( n \b ) ^ { l + l' + 1 } } C _ {l l'} ^ { m } Y _ { l' m } (\t , \varphi) \, e ^ {- |\nu| \b} \\
&=& \f {R _ w ^ { l + 1 } } {r ^ { l + 1 }} Y _ { l m } (\t , \varphi) + \sum _ {l'} \f {R _ w ^ { l' } r ^ { l + 1 }}
{ \b  ^ { l + l' + 1 } } \zeta (l + k + 1) C _ {l l'} ^ { m } Y _ { l' m } (\t , \varphi) \, \Li _ {l + l' + 1} (e ^ {- |\nu| \b}).
\label{zexpwl}
\eea
This modification can be accounted for by the substitution $\zeta (l+l'+1) \to \Li_{l+l'+1}(e^{-|\nu|/T})$. Thus, the catalysis factor becomes (with $d-3$ transverse dimensions)
\be
\mathcal { K } _ w = \det \l [  1 + \bar { \mathcal { D } } ( T , \nu) \r ] ^ { - \f {d-3} {2} },
\ee
where the matrix $\bar { \mathcal { D } } ( T , \nu)$ has a block diagonal form with elements
\be
[{\cal D} _ m] _ { l l'} ( T , \nu) = \f {l \, (l-1)} {(l+1) \, (l+2)} \, C _ { l l'} ^ {m} \l ( {R} {T} \r ) ^ { l + l' +1} \, \Li_{l+l'+1}(e^{-|\nu|/T}).
\label{block3d}
\ee

In order to separate the contribution of binary collisions we expand the catalysis factor in powers of ${\cal D}$ to the second order:
\be
\mathcal { K }_w = 1+ \f {d-3} {4} \Tr \, \bar { \mathcal { D } } ^ 2 ( T , \nu) =
\f {d-3} {4} \sum _ m \Tr \, \mathcal {D} ^ 2_ m,
\label{kfwlt}
\ee
where we have taken into account that the linear term vanishes:
\be
\Tr \, \bar { \mathcal { D } } ( T , \nu) = \sum _ m \Tr \, \mathcal {D} _ m = 0,
\ee
due to the relation (\ref {restr}). Similarly to the case of string decay the physical reason for the absence of the first power is the Lorentz invariance of the system (there is no destruction of the wall by presence of one massless particle). It should be also noted that unlike in the decay of a string the destruction of a domain wall can occur in collisions of an odd number of particles, since the expansion (\ref{kfwlt}) generally contains any larger then one power of matrix $\bar {\mathcal {D}}$.

For an actual calculation of binary processes one can now follow the same steps as in the calculation of the section \ref{sec_therm_bath}. The only difference from the (1+1) dimensional string geometry is that
now the kinematical invariant $s$ also depends on the angle $\t$ between the two particles' momenta,
\be
s (\vec {k} _ 1, \vec {k} _ 2) = 2 \, \omega _ 1 \, \omega _ 2 \, ( 1 + \cos \t )~.
\ee
Therefore an analog of the integral (\ref{bosint}) has an angular part, and one should make use of the relation
\be
\int n( \vec {k} _ 1 ) \, n( \vec {k} _ 2 ) s ^ { n - 1 } (\vec {k} _ 1, \vec {k} _ 2) \f { d ^ 2 k _ 1 \, d ^ 2 k _ 2 } { (2 \pi)^4 } = \f {4 ^ {n - 1}} {( 2 \pi ) ^ 3} \, (n!) ^ 2 \, T ^ { 2 (n + 1) } \, \Li _ {n+1} ^ 2 \l (e ^ { - {|\nu|} / {T} } \r ) \, B (n-1/2,1/2),
\label{bosint2}
\ee
where $B(a,b)$ is the Euler beta function
\be
B ( a, b ) = \f {\Gamma (a) \Gamma (b)} { \Gamma (a+b)}.
\ee
The appropriate quantity describing the probability of a binary process on a (2+1) dimensional domain wall is the effective length $\lambda$, which is an analog of the cross section in (3+1) dimensions and of the dimensionless probability $W_2$ on a (1+1) dimensional string.

Using the expression (\ref{bosint2}) and the quoted in Eq.(\ref{kfwl}) low temperature behavior of the thermal catalysis factor one readily finds the low energy limit for the effective length of destruction of the wall in a collision of two Goldstone bosons:
\be
\lambda = \f {d-3} {5} \, \pi ^ 2 \, \gamma _w \, s ^ 3 \, R _ w ^ {10} + \dots ~.
\ee

The general formula for the effective length at arbitrary values of $\sqrt { s } \, R_ w$ is found using Eq.(\ref{kfwlt}) and the expression (\ref{block3d}) for the elements of the matrix $\tilde {\cal D}$. In this way we arrive at the following expression for the effective length of a two particle collision in the form of a triple sum
\bea
\lambda &=& \sum _ m \lambda _ m =  \f { 16 \pi ^ 3 \, (d-3) \, R _ w ^2 \gamma_w   } {s} \sum _ m \sum _ { p , q \geq |m|}
\f {q \, ( q - 1 ) \, p \, ( p - 1 )} {(q+1) \, (q+2) \, (p+1) \, (p+2)} \nonumber \\ 
&\times& \l ( \f { R _ w \sqrt {s}} {2} \r ) ^ { 2 p +2 q } \, B ^ {-1} (p + q - 1/2, 1/2) \,
\f {1} {(q+m)! \, (q-m)! \, (p+m)! \, (p-m)!}\, ,
\eea
where $\gamma _ w$ is a decay rate per unit area of the wall at zero temperature.

The behavior at large energy, $R_w \, \sqrt{s} \gg 1$, can be found using saddle point approximation to estimate the sum over $p$ and $q$, which gives
\be
\lambda \sim \exp \l ( - \f { 16 \, \pi \, \s ^ 3 } { 3 \, \ep ^2 } + 4 \sqrt {s} \, \f { \s } { \ep } \r ),
\ee
which matches the semiclassical expression for the tunneling exponent at energy $\sqrt{s}$ \cite{v1994}.

%\newpage
\section{Summary and conclusions}
In this paper we considered the decays of metastable topological configurations such as string and domain wall. For each process we found the rate at zero temperature and calculated catalysis factor at finite temperature. Using the relation between catalysis factor and probability (effective) length of a collision of the Goldstone bosons we found probability of a string (domain wall) decay in a collision of two particles. The main conclusions that can be drawn from the presented here study can be formulated as follows
\renewcommand{\theenumi}{\roman{enumi}}
\begin{enumerate}
\item{The effect on the preexponential factor in the tunneling decay rate arising from the Goldstone modes living on strings and walls is expressed in terms of boundary terms in the path integral around the bounce configuration. This effect is fully calculable within the low-energy effective action of the Nambu-Goto type, and essentially reduces to a renormalization of the tension of the interface between the two phases of the topological defect. We have found that this property is unique for the strings and walls and is generally lost in a description of decay of topological objects with more dimensions, where one would have to go beyond the effective Nambu-Goto description.}
\item{The thermal effects in the tunneling rate are also fully calculable and are described by an expansion in powers of the temperature. The power-like, rather than exponential, behavior of the thermal terms is due to the presence of the massless Goldstone bosons.}
\item{The expansion for the thermal catalysis factor converges as long as the temperature is less than the inverse diameter of the bounce, so that in the Euclidean calculation of the bounce configuration to (the  imaginary part of) the free energy there is no obstruction to periodically replicating the bounce with the Matsubara period $\beta=1/T$. For the decay of strings this range of temperatures corresponds to the dominance of the (thermally enhanced) quantum tunneling over the classical thermal effects, and at the critical temperature $1/\ell_c$ the regime of the transition changes and the classical thermal effects take over. For the domain wall decay the classical thermal activation starts dominating at a lower temperature, where the thermal enhancement of the quantum tunneling is still very small.}
\item{The thermal enhancement of the tunneling rate can be viewed as and additional contribution to the decay probability arising from collisions of the Goldstone bosons which are present in the thermal state. Such an interpretation in fact allows to determine the rate of destruction of the considered topological objects in collisions of the Goldstone bosons. In order to unambiguously identify in the thermal expansion the terms corresponding to the contribution of collisions between a given number of the bosons, we have modified the considered statistical ensemble by introducing a (fictitious) negative chemical potential $\nu$ for the Goldstone bosons. The calculated dependence on $\nu$ and $T$ of the preexponential factor in the decay rate thus serves as a generating function for the rates of the decay induced by collisions of particles, including the dependence of those rates on the energies in the collisions.}
\item{We find that a destruction of a string takes place only in collisions of even number of the Goldstone bosons, while a metastable wall can be destroyed in collisions of any number of the bosons, $n \ge 2$.}
\item{We have calculated in a closed form the energy behavior of the probability of destruction of a string in a collision of two Goldstone bosons, which is valid at an arbitrary relation between the energy $\sqrt{s}$ and the infrared scale in the problem $R_s$. The found expression takes an especially simple form of Eq.(\ref{w20}) for the decay of a string into `nothing'}
\item{At large values of $R \,\sqrt{s} $ the collision-induced decay rate contains an exponential factor $\exp(2 R \,\sqrt{s})$, which has been previously argued\cite{vs,v1994} within the leading semiclassical approach and corresponds to a full transfer of the energy from the colliding particles to the tunneling degrees of freedom.}
\end{enumerate}

One can readily notice that our calculation of the collision-induced rate through a thermal ensemble with an artificially introduced chemical potential is quite indirect. In a sense, this approach is reminiscent of the treatment in Ref.~\cite{Luscher}, where finite volume effects in the energy of a two-particle state are related to the binary scattering amplitude. The obvious difference is that we calculate the free energy of a statistical ensemble rather than of a particular quantum state. In relation to this part of our calculation it would certainly be illuminating to have a more direct, and possibly simpler method for calculating the probability of creating semiclassical objects, such as the bubbles of the stable phase, in collisions of particle. However, lacking such a method at present, we have to resort to the indirect calculation described in the present paper. 

\section*{Acknowledgments}
This work is supported  in part by the DOE grant DE-FG02-94ER40823. The work of
A.M. is also supported in part by the RFBR Grant No. 07-02-00878 and by the
Scientific School Grant No. NSh-3036.2008.2.

%\newpage

\end{document}